\documentclass[twocolumn,showpacs,aps,prd,floatfix,preprintnumbers]{revtex4}
\usepackage{graphicx} 
\usepackage{dcolumn}
\usepackage{epsfig}    
\usepackage{amsmath}
\usepackage{color}
\input babarsym   

\newcommand{\kskpi}{\ensuremath{\KS\Kpm \pimp}\xspace}

\renewcommand{\mm}{\ensuremath{M^2_{\mathrm{rec}}}\xspace}

\newcommand{\kkpiz}{\ensuremath{\Kp\Km \piz}\xspace}
\newcommand{\etacpiz}{\ensuremath{\etac \to \Kp\Km \piz}\xspace}
\newcommand{\etactokkpi}{\ensuremath{\etac \to \KS\Kpm \pimp}\xspace}
\def\Kstarz   {\ensuremath{K^*_0(1430)^0}\xspace}
\def\Kstarzp   {\ensuremath{K^*_0(1430)^+}\xspace}
\def\Kstarzero   {\ensuremath{K^*_0(1430)}\xspace}

\def\Kstarzp   {\ensuremath{K^*_0(1950)}\xspace}

\def\Kstarone {\ensuremath{K^*(892)}\xspace}

\renewcommand{\gg}{\ensuremath{\gamma\gamma}}

\mathchardef\myhyphen="2D
\newcommand{\BaBarPubYear}    {15}
\newcommand{\BaBarPubNumber}  {008}
\newcommand{\SLACPubNumber}   {16422}

\begin{document}
\begin{flushleft}
\babar-PUB-\BaBarPubYear/\BaBarPubNumber \\
SLAC-PUB-\SLACPubNumber \\
\end{flushleft}

\title{
 \large \bf\boldmath Measurement of the $I=1/2$  $K \pi$ $\mathcal{S}$-wave amplitude from Dalitz plot analyses of $\eta_c \to K \Kbar \pi$ in
 two-photon interactions
}

%
%
\author{J.~P.~Lees}
\author{V.~Poireau}
\author{V.~Tisserand}
\affiliation{Laboratoire d'Annecy-le-Vieux de Physique des Particules (LAPP), Universit\'e de Savoie, CNRS/IN2P3,  F-74941 Annecy-Le-Vieux, France}
\author{E.~Grauges}
\affiliation{Universitat de Barcelona, Facultat de Fisica, Departament ECM, E-08028 Barcelona, Spain }
\author{A.~Palano}\altaffiliation{Also at: Thomas Jefferson National Accelerator Facility,  Newport News, Virginia 23606, USA}
\affiliation{INFN Sezione di Bari and Dipartimento di Fisica, Universit\`a di Bari, I-70126 Bari, Italy }
\author{G.~Eigen}
\affiliation{University of Bergen, Institute of Physics, N-5007 Bergen, Norway }
\author{D.~N.~Brown}
\author{Yu.~G.~Kolomensky}
\affiliation{Lawrence Berkeley National Laboratory and University of California, Berkeley, California 94720, USA }
\author{H.~Koch}
\author{T.~Schroeder}
\affiliation{Ruhr Universit\"at Bochum, Institut f\"ur Experimentalphysik 1, D-44780 Bochum, Germany }
% ...
\collaboration{The \babar\ Collaboration}
\noaffiliation

\begin{abstract}
We study the processes $\gg\to \kskpi$ and $\gg\to \kkpiz$ using a data sample
of 519~\invfb\ recorded with the \babar\ detector operating at the SLAC PEP-II
asymmetric-energy \epem\ collider at center-of-mass energies at and near the
$\Upsilon(nS)$ ($n = 2,3,4$) resonances.
We observe \etac decays to both final states and perform Dalitz plot analyses
using a model-independent partial wave analysis technique.
This allows a model-independent measurement of the mass-dependence of the $I=1/2$  $K \pi$ $\mathcal{S}$-wave amplitude and phase.
A comparison between the present measurement and those from previous experiments indicates similar behaviour for the phase up to a mass of
1.5 \gevcc. In contrast, the amplitudes show very marked differences.
The data require the presence of a new $a_0(1950)$ resonance with parameters
$m=1931 \pm 14 \pm 22 \ \mevcc$ and $\Gamma=271 \pm 22 \pm 29 \ \mev$.
\end{abstract}
\maketitle

\section{Introduction}
Scalar mesons are still a puzzle in light meson spectroscopy: they have complex structure,
and there are too many states to be accommodated within the quark model without difficulty~\cite{polosa}. 
In particular, the structure of the $I=1/2$  $K \pi$ $\mathcal{S}$-wave is a longstanding problem. In recent years many experiments have performed 
accurate studies of the decays of heavy-flavored hadrons producing a $K \pi$ system in the final state.
These studies include searches for \CP violation~\cite{cp}, and searches for, and observation of, new exotic resonances~\cite{zs} and
charmed mesons~\cite{bs}.
However, the still poorly known structure of 
the  $I=1/2$  $K \pi$ $\mathcal{S}$-wave is a source of large systematic uncertainties.
The best source of information on the scalar structure of the $K \pi$ system comes from the LASS experiment, which studied the reaction $\Km p \to \Km \pip n$~\cite{lass_kpi}.
Partial wave analysis of the $K \pi$ system reveals a large contribution from the $I=1/2$  $K \pi$ $\mathcal{S}$-wave amplitude over the mass
range studied.
In the description of the $I=1/2$  scalar amplitude up to a $K \pi$ mass of about 1.5 \gevcc\ the $K^*_0(1430)$ resonant amplitude is added coherently to an effective-range
description in such a way that the net amplitude
actually decreases rapidly at the resonance mass. The $I=1/2$  $\mathcal{S}$-wave amplitude representation is given explicitly in Ref.~\cite{babar_z}.
In the LASS analysis, in the region above 1.82 \gevcc, the $\mathcal{S}$-wave suffers from a two-fold ambiguity, but in both solutions it is understood in terms of the presence of a $K^*_0(1950)$ resonance. It should be noted that the extraction of the $I=1/2$ $\mathcal{S}$-wave amplitude is complicated by the presence of an $I=3/2$ contribution. 

Further information on the $K \pi$ system has been extracted from Dalitz plot analysis of the decay $D^+ \to \Km \pip \pip$ where, in order to fit 
the data, the presence of an additional resonance, the $\kappa(800)$, was claimed~\cite{aitala}. Using the same data, a Model Independent Partial Wave Analysis (MIPWA)
of the $K \pi$ system was developed for the first time~\cite{aitala1}.
This method allows the amplitude and phase of the $K \pi$ $\mathcal{S}$-wave to be extracted as
 functions of mass (see also Refs.~\cite{cleo} and ~\cite{focus}). However in these analyses the phase space is limited to mass values less than 1.6 \gevcc\ due to the kinematical limit imposed by the $D^+$ mass.
A similar method has been used to extract the $\pip \pim$ $\mathcal{S}$-wave amplitude in a Dalitz plot analysis of $D^+_s \to \pip \pim \pip$~\cite{marco}.

In the present analysis, we consider three-body \etac decays to $K \Kbar \pi$ and obtain new information on the $K \pi$ $I=1/2$  $\mathcal{S}$-wave amplitude extending up to a mass of 2.5 \gevcc. We emphasize that, due to isospin conservation in the \etac hadronic decay to $(K \pi) \Kbar$,
the $(K \pi)$ amplitude must have $I=1/2$ , and there is no $I=3/2$ contribution.
The \babar\ experiment first performed a Dalitz plot analysis of $\etac \to \Kp \Km \piz$ and $\etac \to \Kp \Km \eta$ using an isobar model~\cite{etakk}. The analysis reported the first observation of $K^*_0(1430) \to K \eta$, and observed that \etac decays to three pseudoscalars are dominated by intermediate
scalar mesons. A previous search for charmonium resonances decaying to \kskpi in two-photon interactions is reported in Ref.~\cite{kkpipipi0}. 
We continue these studies of \etac decays and extract the $K \pi$ $\mathcal{S}$-wave amplitude by performing a MIPWA of both \etactokkpi and \etacpiz final states. 

We describe herein studies of the $K \Kbar \pi$ system produced in two-photon interactions.
Two-photon events in which at least one of the interacting photons is not quasi-real are
strongly suppressed by the selection 
criteria described below. This implies that the allowed $J^{PC}$ values of
any produced resonances are $0^{\pm+}$, $2^{\pm+}$, $3^{++}$, $4^{\pm+}$...~\cite{Yang}.       
Angular momentum conservation, parity conservation, and charge conjugation
invariance imply that these quantum numbers also apply to
the final state except that the $K \Kbar \pi$ state cannot be in a $J^P=0^+$ state.

This article is organized as follows. In Sec.\ II, a brief description of the
\babar\ detector is given. Section III is devoted to the event reconstruction and data selection of the \kskpi system. In Sec.\ IV, we describe studies of efficiency and resolution,
while in Sec.\ V we describe the MIPWA. In Secs. VI and VII we perform Dalitz plot analyses of \etactokkpi and \etacpiz decays. Section VIII is devoted to discussion of the measured $K \pi$ $\mathcal{S}$-wave amplitude, and finally results are summarized in Sec.~IX.

\section{The \babar\ detector and dataset}

The results presented here are based on data collected
with the \babar\ detector 
at the PEP-II asymmetric-energy $e^+e^-$ collider
located at SLAC, and correspond 
to an integrated luminosity of 519~\invfb~\cite{luminosity} recorded at
center-of-mass energies at and near the $\Upsilon (nS)$ ($n=2,3,4$)
resonances. 
The \babar\ detector is described in detail elsewhere~\cite{BABARNIM}.
Charged particles are detected, and their
momenta are measured, by means of a five-layer, double-sided microstrip detector,
and a 40-layer drift chamber, both operating  in the 1.5~T magnetic 
field of a superconducting
solenoid. 
Photons are measured and electrons are identified in a CsI(Tl) crystal
electromagnetic calorimeter. Charged-particle
identification is provided by the measurement of specific energy loss in
the tracking devices, and by an internally reflecting, ring-imaging
Cherenkov detector. Muons and \KL\ mesons are detected in the
instrumented flux return  of the magnet.
Monte Carlo (MC) simulated events~\cite{geant}, with reconstructed sample sizes 
more than 10  times larger than the corresponding data samples, are
used to evaluate the signal efficiency and to determine background features. 
Two-photon events are simulated  using the GamGam MC
generator~\cite{BabarZ}.

\section{ {\boldmath Reconstruction and selection of $\protect \etactokkpi$} events}

To study the reaction
\begin{equation}
\gamma \gamma \to \KS \Kpm \pimp
\end{equation}
we select events in which the $e^+$ and $e^-$  beam particles are scattered  
at small angles, and hence are undetected in the final state. 
We consider only events for which the number of well-measured charged-particle tracks with
transverse momentum greater than 0.1~\gevc\ is exactly equal to four, and for which there are no more than five photon candidates
with reconstructed energy in the electromagnetic calorimeter greater than 100 MeV.
We obtain $K^0_S \to \pip \pim$ candidates by means of a vertex fit of pairs of oppositely charged tracks which
requires a $\chi^2$ fit probability greater than 0.001. Each \KS candidate is then combined with 
two oppositely charged tracks, and fitted to a common vertex, with the requirements that the fitted vertex be within the
$e^+ e^-$ interaction region and have a $\chi^2$ fit probability greater than 0.001.
We select kaons and pions by applying high-efficiency particle identification criteria.
We do not apply any particle identification requirements
to the pions from the \KS decay.
We accept only $K_S^0$ candidates with decay lengths from the main vertex of the event greater than 0.2 cm, and
require $\cos \theta_{\KS}>0.98$, where $\theta_{\KS}$ is defined as the angle between the \KS momentum direction and the
line joining the primary and the \KS vertex.
A fit to the $\pip \pim$ mass spectrum using a linear function for the background and a Gaussian
function with mean $m$ and width $\sigma$ gives $m=497.24$ \mevcc\ and $\sigma=2.9$ \mevcc. We select the $\KS$ signal region to be within
$\pm 2 \sigma$ of $m$ and reconstruct the \KS 4-vector by adding the three-momenta of the pions and computing the energy using the \KS PDG mass value~\cite{pdg}.

\begin{figure}
\begin{center}
\includegraphics[width=9cm]{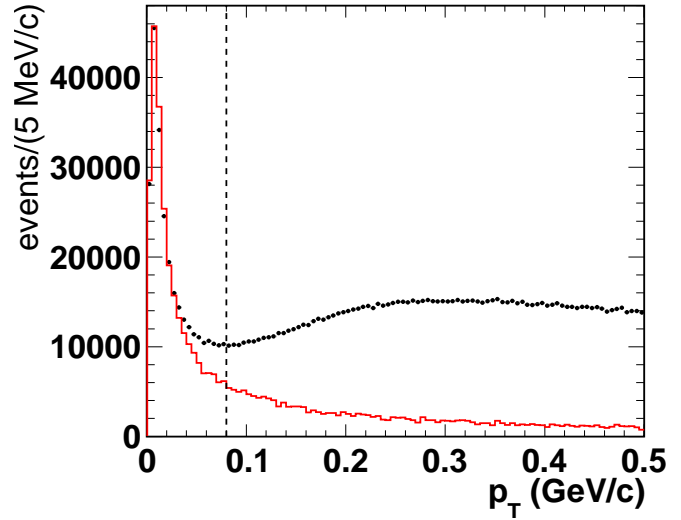}
\caption{Distributions of \pt\ for $\gamma \gamma \to \KS \Kpm \pimp$.  The data are shown as (black) points with error bars,
and the signal MC simulation as a (red) histogram; the vertical dashed line indicates the selection applied to select two-photon events.}
\label{fig:fig1}
\end{center}
\end{figure}
Background arises mainly from random combinations of particles from
\epem\ annihilation, from other two-photon processes, and from events with initial-state photon radiation (ISR). The ISR 
background is dominated by $J^{PC}=1^{--}$ resonance production~\cite{isr}.
We discriminate against \kskpi events produced via ISR by requiring $\mm\equiv(p_{\epem}-p_{\mathrm{rec}})^2>10$~GeV$^2$/$c^4$, where
$p_{\epem}$ is the four-momentum of the initial state and $p_{\mathrm{rec}}$ is the four-momentum of the \kskpi system. 

The \kskpi mass spectrum shows a prominent \etac signal.
We define \pt\ as the magnitude of the vector sum of the transverse momenta, in the \epem\ rest frame, of the final-state particles with respect to the beam axis.
Since well-reconstructed two-photon events are expected to have low values of \pt, we optimize the selection as a function of this variable.
We produce $\kskpi$ mass spectra with different \pt selections and fit the mass spectra to extract the number of \etac signal events ($N_s$) and the number
of background events below the \etac signal ($N_b$). We then compute the purity, defined as $P = N_s/(N_s + N_b)$, and the significance $S = N_s/\sqrt{N_s + N_b}$. To obtain the best significance with the highest purity, 
we optimize the selection by requiring  the maximum value of the product of purity and significance, $P \cdot S$, and find that this corresponds to the requirement $\pt<0.08~\gevc$.

\begin{figure}
\begin{center}
\includegraphics[width=9cm]{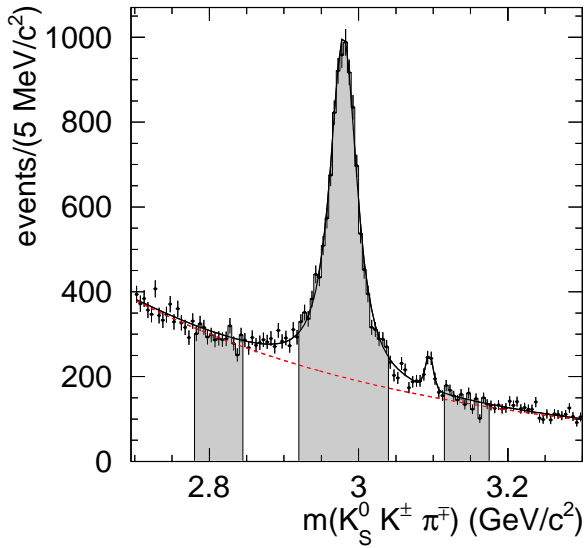}
\caption{The $\KS \Kpm \pimp$ mass spectrum in the \etac mass region after requiring $\pt<0.08~\gevc$. The solid curve shows the total fitted function,
and the dashed curve shows the fitted background contribution. The shaded areas show signal and sideband regions.}
\label{fig:fig2}
\end{center}
\end{figure}

Figure~\ref{fig:fig1} shows the  measured \pt\ distribution in comparison to the corresponding  \pt\ distribution obtained from simulation of the signal process.
A peak at low \pt\ is observed indicating
the presence of the two-photon process. The shape of the peak agrees well with that seen in the MC simulation. 
Figure~\ref{fig:fig2} shows the $\KS \Kpm \pimp$ mass spectrum in the \etac mass region. A clear \etac signal over a background of about 35\% can be seen, together with a residual \jpsi signal. Information on the fitting procedure is given at the end of Sec. IV.
We define the \etac signal region as the range 2.922-3.039~\gevcc\ ($m(\etac) \pm 1.5 \ \Gamma$), which contains 12849 events with a purity of
$(64.3 \pm 0.4)$\% . 
Sideband regions are defined by the ranges 2.785-2.844~\gevcc \ and 3.117-3.175~\gevcc\ (within 3.5-5 $\Gamma$), respectively as indicated (shaded) in
Fig.~\ref{fig:fig2}.

Details on data selection, event reconstruction, resolution, and efficiency measurement for the \etacpiz decay can be found in Ref.~\cite{etakk}.
The \etac signal region for this decay mode contains 6710 events with a purity of $(55.2 \pm 0.6)$\%. 

\section{Efficiency and resolution}

To compute the efficiency, MC signal events are generated using a detailed detector simulation~\cite{geant} in which the \etac decays uniformly in phase space.
These simulated events are reconstructed and analyzed in the same manner as data. The efficiency is computed as the ratio of 
reconstructed to generated events. 
Due to the presence of long tails in the Breit-Wigner (BW) representation of the resonance, we apply 
selection criteria to restrict the generated events to the \etac mass region. 
We express the efficiency as a function of the invariant mass, $m(\Kp \pim)$~\cite{conj}, and $\cos \theta$, where $\theta$ is the angle, in the $\Kp \pim$ 
rest frame, between the directions of the \Kp\ and the boost from the $\KS \Kp \pim$ rest frame.

\begin{figure}
\begin{center}
\includegraphics[width=9cm]{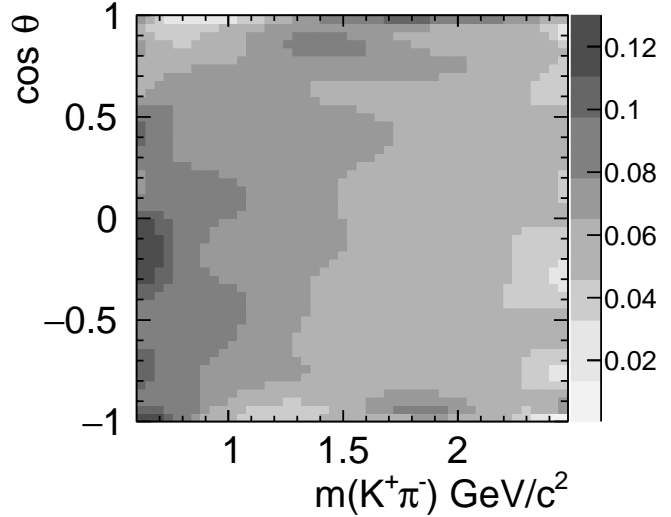}
\caption{Fitted detection efficiency in the $\cos \theta \  vs. \ m(\Kp \pim)$ plane. Each interval shows the average value of the fit for that region.}
\label{fig:fig3}
\end{center}
\end{figure}

To smooth statistical fluctuations, this efficiency map is parametrized as follows.
First we fit the efficiency as a function of  $\cos \theta$ in separate intervals of $m(\Kp \pim)$, using 
Legendre polynomials up to $L=12$:
\begin{equation}
\epsilon(\cos\theta) = \sum_{L=0}^{12} a_L(m) Y^0_L(\cos\theta),
\end{equation}
where $m$ denotes the $\Kp \pim$ invariant mass.
For each value of $L$, we fit the mass dependent coefficients $a_L(m)$ with a seventh-order polynomial in $m$.
Figure~\ref{fig:fig3} shows the resulting fitted efficiency map $\epsilon(m,\cos \theta)$.
We obtain $\chi^2/N_{\rm cells}=217/300$ for this fit, where $N_{\rm cells}$ is the number of cells in the efficiency map.
We observe a significant decrease in
efficiency in regions of $\cos\theta \sim \pm 1$ due to the impossibility of reconstructing $\Kpm$ mesons with laboratory momentum
less than about 200~\mevc, and \pipm and $\KS(\to \pip \pim)$ mesons with laboratory momentum less than about 100~\mevc (see Fig. 9 of Ref.~\cite{babar_z}). These effects result from energy loss in the beampipe and inner-detector material.

The mass resolution, $\Delta m$, is measured as the difference between the generated and reconstructed \kskpi invariant-mass values.
The distribution has a root-mean-squared value of 10 \mevcc, and is parameterized by the sum of a Crystal Ball~\cite{cb} and a Gaussian function. 
We perform a binned fit to the \kskpi mass spectrum in data using the following model. The background is described by a second-order polynomial, and the \etac resonance is represented by a nonrelativistic BW function convolved with the resolution function. 
In addition, we allow for the presence of a residual \jpsi contribution modeled with a Gaussian function. Its parameter values are fixed to those obtained from a fit to the \kskpi mass spectrum for the ISR data sample obtained by requiring
$\lvert \mm \rvert<1 \ {\rm GeV}^2/c^4$.
The fitted \kskpi mass spectrum is shown in Fig.~\ref{fig:fig2}. We obtain the following \etac parameters:
\begin{equation}
  \begin{split}
    m=2980.8 \pm 0.4 \ \mevcc, \ \Gamma=33 \pm 1 \ \mev,\\
    \ N_{\etac}=9808 \pm 164,
    \end{split}
\end{equation}
where uncertainties are statistical only. Our measured mass value is 2.8 \mevcc\ lower than the world average~\cite{pdg}.
This may be due to interference between the \etac amplitude and that describing the background in the signal region~\cite{bes_int}.

\section{Model Independent Partial Wave Analysis}

We perform independent MIPWA of the \kskpi and \kkpiz Dalitz plots in the \etac mass region using  unbinned maximum likelihood fits.
The likelihood function is written as
\begin{eqnarray}
\mathcal{L} = \nonumber\\     
 \prod_{n=1}^N&\bigg[&f_{\rm sig}(m_n) \epsilon(x'_n,y'_n)\frac{\sum_{i,j} c_i c_j^* A_i(x_n,y_n) A_j^*(x_n,y_n)}{\sum_{i,j} c_i c_j^* I_{A_i A_j^*}} \nonumber\\
& &+(1-f_{\rm sig}(m_n))\frac{\sum_{i} k_iB_i(x_n,y_n,m_n)}{\sum_{i} k_iI_{B_i}}\bigg]
\end{eqnarray}
\noindent where
\begin{itemize}
\item $N$ is the number of events in the signal region;
\item for the $n$-th event, $m_n$ is the \kskpi or the \kkpiz invariant mass;
\item for the $n$-th event, $x_n=m^2(K^+ \pim)$, $y_n=m^2(\KS \pim)$ for \kskpi; $x_n=m^2(K^+ \piz)$, $y_n=m^2(K^- \piz)$ for $\kkpiz$; 
\item $f_{\rm sig}$ is the mass-dependent fraction of signal obtained from the fit to the \kskpi or \kkpiz mass spectrum;
\item for the $n$-th event, $\epsilon(x'_n,y'_n)$ is the efficiency parametrized as a function of $x'_n=m(\Kp \pim)$ for \kskpi and $x'_n=m(\Kp \Km)$ for \kkpiz, and $y'_n=\cos \theta$ (see Sec. IV);
\item for the $n$-th event, the $A_i(x_n,y_n)$ describe the complex signal-amplitude contributions;
\item $c_i$ is the complex amplitude for the $i$-th signal component; the $c_i$ parameters are allowed to vary during the fit process;
\item for the $n$-th event, the $B_i(x_n,y_n)$ describe the background probability-density functions assuming that interference between signal and background amplitudes can be ignored;
\item $k_i$ is the magnitude of the $i$-th background component; the $k_i$ parameters are obtained by fitting the sideband regions;
\item $I_{A_i A_j^*}=\int A_i (x,y)A_j^*(x,y) \epsilon(x', y')\ {\rm d}x{\rm d}y$ and 
$I_{B_i}~=~\int B_i(x,y) {\rm d}x{\rm d}y$ are normalization
  integrals. Numerical integration is performed on phase space generated events with \etac signal and background generated according to the experimental distributions. In case of MIPWA or when resonances have free parameters, integrals are re-computed at each minimization step.
  Background integrals and fits dealing with amplitudes having fixed resonance parameters are computed only once. 
\end{itemize}
Amplitudes are described along the lines described in Ref.~\cite{kopp}.
For an \etac meson decaying into three pseudoscalar mesons via an intermediate
  resonance $r$ of spin $J$ (i.e. $\etac \to C r$, $r \to A B$), each amplitude
  $A_i(x,y)$ is represented by the product of a complex Breit-Wigner (BW)
  function and a real angular distribution function represented by
  the spherical harmonic function $\sqrt{2 \pi} Y_J^0({\rm cos} \theta)$; $\theta$
  is the angle between the direction of $A$, in the rest frame of $r$,
  and the direction of $C$ in the same frame. This form of the angular
  dependence results from angular momentum conservation in the rest
  frame of the \etac, which leads to the production of $r$ with helicity 0.

  It follows that
\begin{equation}
    A_i(x,y) = BW(M_{AB}) \sqrt{2 \pi}  Y_J^0({\rm cos} \theta).
\label{eq:spin}
\end{equation}

  The function $BW(M_{AB})$ is a relativistic BW function of the form
  \begin{equation}
         BW(M_{AB}) = \frac{F_{\eta_c} F}{M_r^2 - M_{AB}^2 - i M_r \Gamma_{\rm tot}(M_{AB})}
\end{equation}
   where $M_r$ is the mass of the resonance $r$, and $\Gamma_{\rm tot}(M_{AB})$ is
   its mass-dependent total width. In general, this mass dependence
   cannot be specified, and a constant value should be used. However,
   for a resonance such as the $K^*_0(1430)$, which is approximately elastic,
   we can use the partial width $\Gamma_{AB}$, and specify the mass-dependence
   as:
\begin{equation}
\Gamma_{AB} = \Gamma_r \left(\frac{p_{AB}}{p_r}\right)^{2J+1} \left(\frac{M_r}{M_{AB}}\right)F^2
\end{equation}
where
\begin{equation}
p_{AB} = \frac{\sqrt{\left(M_{AB}^2-M_A^2-M_B^2\right)^2-4M_A^2M_B^2}}{2M_{AB}}.
\end{equation}
and $p_r$ is the value of $p_{AB}$ when $M_{AB}=M_r$. 

The form factors $F_{\etac}$ and  $F$ attempt to model the underlying quark structure of the parent particle and the intermediate
resonances. We set $F_{\etac}$ to a constant value, while for $F$  we use Blatt-Weisskopf penetration factors~\cite{blatt} (Table~\ref{tab:tab_blatt}), that depend on a single parameter $R$ representing the meson ``radius'', for which we assume $R=1.5 \ \gev^{-1}$.
The $a_0(980)$ resonance is parameterized as a coupled-channel Breit-Wigner function whose parameters are taken from Ref.~\cite{cbar}.

\begin{table}
  \caption{Summary of the Blatt-Weisskopf penetration form factors.}
  \label{tab:tab_blatt}
\begin{center}
  \begin{tabular}{cc}
    \hline \\ [-2.3ex]
Spin & $F$ \\
\hline \\ [-2.3ex]
0 & $1$ \\
&\\
1 & {\Large $\frac{\sqrt{1+(R_r p_r)^2}}{\sqrt{1+(R_r p_{AB})^2}}$} \\
&\\
2 & {\Large $\frac{\sqrt{9+3(R_r p_r)^2+(R_r p_r)^4}}{\sqrt{9+3(R_r p_{AB})^2+(R_r p_{AB})^4}}$} \\
\hline
\end{tabular}
\end{center}
\end{table}

To measure the $I=1/2$  $K \pi$ $\mathcal{S}$-wave we make use of a MIPWA technique first described in Ref.~\cite{aitala1}.
The $K \pi$ $\mathcal{S}$-wave, being the largest contribution, is taken as the reference amplitude. 
We divide the $K \pi$ mass spectrum into 30 equally-spaced mass intervals 60 \mev\ wide, and 
for each interval we add to the fit two new free parameters,
the amplitude and the phase of the $K \pi$ $\mathcal{S}$-wave in that interval. 
We fix the amplitude to 1.0 and its phase to $\pi/2$ at an arbitrary point in the mass spectrum, for which we choose interval 14, corresponding to a mass of 1.45 \gevcc. The number of associated free parameters is therefore 58.

Due to isospin conservation in the hadronic $\eta_c$  and $K^*$ decays, the $(K \pi) \Kbar$ amplitudes are combined with positive signs, and so 
therefore are symmetrized with respect to the two $K^* \Kbar$ modes. In particular we write the $K \pi$ $\mathcal{S}$-wave amplitudes as

\begin{equation}
  A_{\mathcal{S}\myhyphen\rm{wave}} = \frac{1}{\sqrt{2}}(a_j^{K^+ \pi^-}e^{i\phi_j^{K^+ \pi^-}} + a_j^{\Kbar^0 \pi^-}e^{i\phi_j^{\Kbar^0 \pi^-}}),
  \label{eq:amp}
\end{equation}

\noindent where $a^{K^+ \pi^-}(m)=a^{\Kbar^0 \pi^-}(m)$ and $\phi^{K^+ \pi^-}(m) = \phi^{\Kbar^0 \pi^-}(m)$, for $\eta_c \to \Kbar^0 \Kp \pim$~\cite{conj} and

\begin{equation}
  A_{\mathcal{S}\myhyphen\rm{wave}} = \frac{1}{\sqrt{2}}(a_j^{K^+ \pi^0}e^{i\phi_j^{K^+\pi^0}} + a_j^{K^- \pi^0}e^{i\phi_j^{K^- \pi^0}}),
\end{equation}
where $a^{K^+ \pi^0}(m)=a^{K^- \pi^0}(m)$ and $\phi^{K^+ \pi^0}(m) = \phi^{K^- \pi^0}(m)$, for \etacpiz.
For both decay modes the bachelor kaon is in an orbital $\mathcal{S}$-wave with respect to the relevant $K \pi$ system, and so does not affect
these amplitudes. The second amplitude in Eq.(\ref{eq:amp}) is reduced because the $\Kbar^0$ is observed as a $\KS$, but the same reduction
factor applies to the first amplitude through the bachelor $\Kbar^0$, so that the equality of the three-body amplitudes is preserved.

Other resonance contributions are described as above. The $K^*_2(1430) \Kbar$ contribution is symmetrized in the same way as the $\mathcal{S}$-wave amplitude.

We perform MC simulations to test the ability of the method to find the correct solution.
We generate \etactokkpi event samples which yield reconstructed samples having the same size as the data sample, according to arbitrary mixtures of resonances, and extract the $K \pi$ $\mathcal{S}$-wave using the MIPWA method. We find that the fit is able to extract correctly
the mass dependence of the amplitude and phase.

We also test the possibility of multiple solutions by starting the fit from random values or
constant parameter values very far from the solution found by the fit.
We find  only one solution in both final states and conclude that the fit converges to give the correct $\mathcal{S}$-wave behaviour for
different starting values of the parameters.

The efficiency-corrected fractional contribution $f_i$ due to resonant or non-resonant contribution $i$ is defined as follows:
\begin{equation}
f_i = \frac {|c_i|^2 \int |A_i(x_n,y_n)|^2 {\rm d}x {\rm d}y}
{\int |\sum_j c_j A_j(x,y)|^2 {\rm d}x {\rm d}y}.
\end{equation}
The $f_i$ do not necessarily sum to 100\% because of interference effects. The uncertainty for each $f_i$ is evaluated by propagating the full covariance matrix obtained from the fit.

We test the quality of the fit by examining a large sample of MC events at the generator level weighted 
by the likelihood fitting function and by the efficiency.
These events are used to
compare the fit result to the Dalitz plot and its projections with proper normalization.
In these MC simulations we smooth the fitted $K \pi$ $\mathcal{S}$-wave amplitude and phase by means of a cubic spline.
We make use of these weighted events to compute a \mbox{2-D} $\chi^2$ over the Dalitz plot. For this purpose, we divide the Dalitz plot into a grid of $25 \times 25$
cells and consider only those containing at least five events. We compute 
$\chi^2 = \sum_{i=1}^{N_{\rm cells}} (N^i_{\rm obs}-N^i_{\rm exp})^2/N^i_{\rm exp}$, where $N^i_{\rm obs}$ and $N^i_{\rm exp}$ are event yields from data and simulation, respectively.

\begin{figure*}
\begin{center}
\includegraphics[width=12cm]{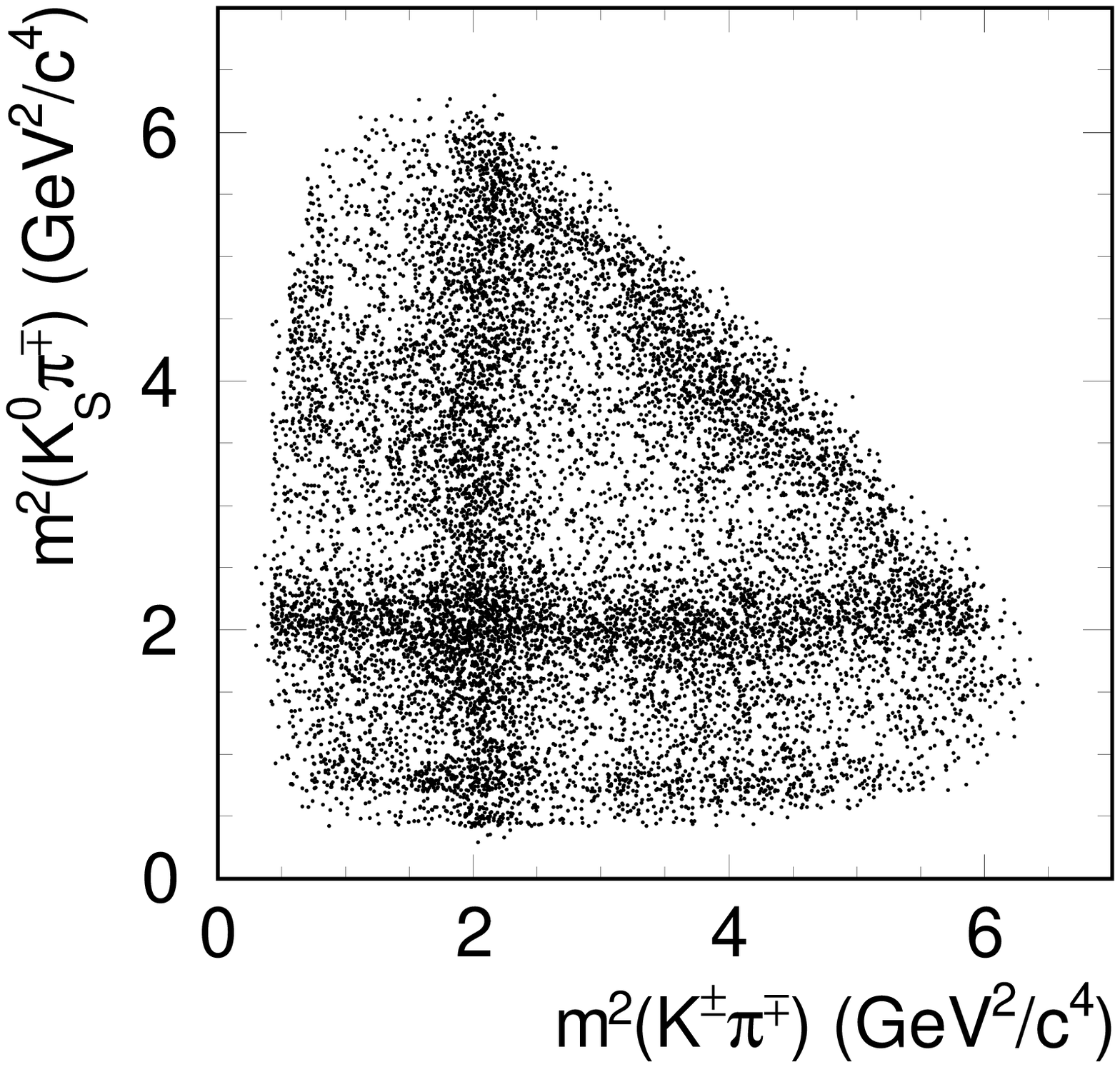}
\caption{Dalitz plot for $\etac \to \kskpi$ events in the signal region.}
\label{fig:fig4}
\end{center}
\end{figure*}

\begin{figure*}
\begin{center}
\includegraphics[width=18cm]{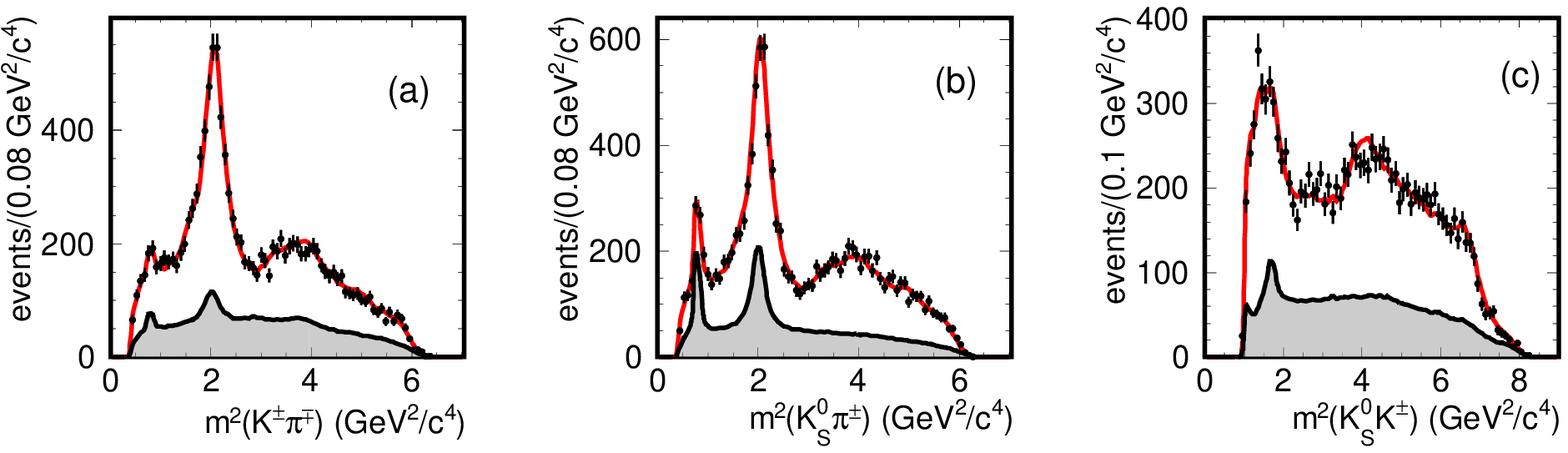}
\caption{The $\etac \to \kskpi$ Dalitz plot projections on (a) $m^2(\Kpm \pimp)$, (b) $m^2(\KS \pipm)$, and (c) $m^2(\KS \Kpm)$. The superimposed curves result from the MIPWA described in the text. The shaded regions show the
background estimates obtained  by interpolating the results of the Dalitz plot analyses of the sideband regions.}
\label{fig:fig5}
\end{center}
\end{figure*}

\section{Dalitz plot analysis of {\boldmath$\protect \etactokkpi$} }

Figure~\ref{fig:fig4} shows the Dalitz plot for the candidates in the \etac signal region, and Fig.~\ref{fig:fig5} shows the corresponding Dalitz plot projections. Since the width of the \etac meson is $32.3 \pm 1.0$ MeV, no mass constraint can be applied.

The Dalitz plot is dominated by the presence of horizontal and vertical uniform bands at the position of the \Kstarzero resonance. We also observe
further bands along the diagonal. Isospin conservation in \etac decay requires that the $(K \Kbar)$ system have I=1, so that these structures
may indicate the presence of $a_0$ or $a_2$ resonances. Further narrow bands are observed at the position
of the $K^*(892)$ resonance, mostly in the $\KS \pim$ projection; these components are consistent with originating from background, as will be shown.

\begin{figure*}
\begin{center}
\includegraphics[width=16cm]{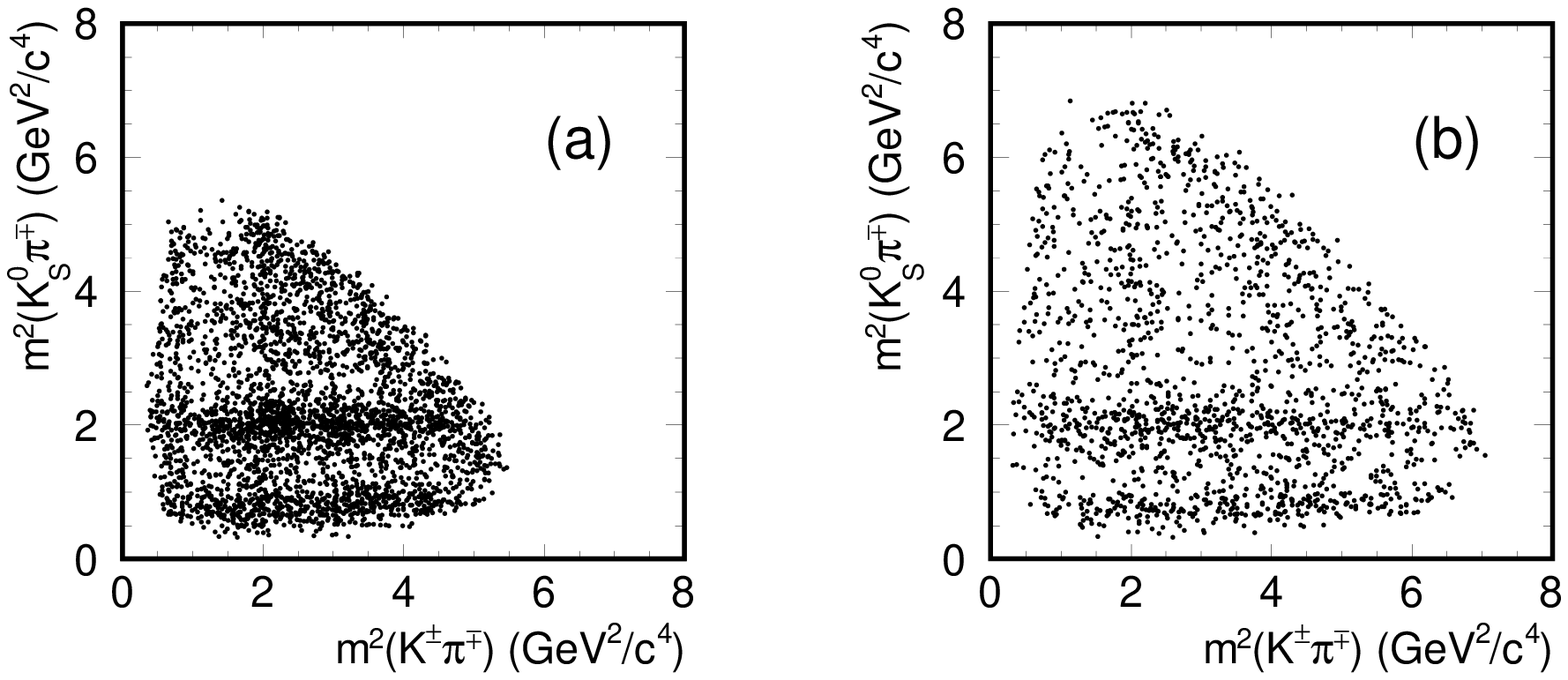}
\caption{Dalitz plots for the $\etac \to \kskpi$ sideband regions: (a) lower, (b) upper.}
\label{fig:fig6}
\end{center}
\end{figure*}

The presence of background in the \etac signal region requires precise study of its structure. This can be achieved by
means of the data in the \etac sideband regions, for which the Dalitz plots are shown in Fig.~\ref{fig:fig6}.

In both regions we observe almost uniformly populated resonant structures mostly in the $\KS \pim$ mass, especially in the regions corresponding to the $K^*(892)$`<
and $K^*_2(1430)$ resonances. The resonant structures in $\Kp \pim$ mass are weaker. The three-body decay of a pseudoscalar meson into a spin-one or spin-two resonance yields a non-uniform distribution (see Eq.~\ref{eq:spin}) in the relevant resonance band on the Dalitz plot. The presence of uniformly populated bands in the $K^*(892)$ and $K^*_2(1430)$ mass regions, indicates that these
structures are associated with background. Also, the asymmetry between the two $K^*$ modes in background 
may be explained as being due to interference between the $I = 0$ and $I = 1$ isospin configurations for the $K^*(\to K \pi) \Kbar$ final
state produced in two-photon fusion.  

We fit the \etac sidebands using an incoherent sum of amplitudes, which includes contributions from the $a_0(980)$, $a_0(1450)$, $a_2(1320)$, $K^*(892)$, $K^*_0(1430)$, $K^*_2(1430)$, $K^*(1680)$, and $K^*_0(1950)$ resonances. To better constrain the sum of the fractions to one, we make use of the channel likelihood method~\cite{chafit} and include resonances
until no structure is left in the background and an accurate description of the Dalitz plots is obtained.

To estimate the background composition in the \etac signal region we perform a linear mass dependent interpolation of the
fractions of the different contributions, obtained from the fits to the sidebands, and normalized using the results from the fit to the \kskpi mass spectrum. The estimated background contributions are indicated by the shaded regions in Fig.~\ref{fig:fig5}.

\subsection{MIPWA of {\boldmath$\protect \etactokkpi$}}

We perform the MIPWA including the resonances listed in Table~\ref{tab:tab1}. In this table, and in the remainder of the paper,
we use the notation $(K \pi) \Kbar$ or $K^* \Kbar$ to represent the corresponding symmetrized amplitude.
After the solution is found we test for
other contributions, including spin-one resonances, but these are found to be consistent with zero, and so are not included.
This supports the observation that
the observed \Kstarone structures originate entirely from background.
We find a dominance of the $K \pi$ $\mathcal{S}$-wave amplitude, with small contributions from $a_0 \pi$ amplitudes and a
significant $K^*_2(1430) \Kbar$ contribution.

\begin{table*}
\caption{Results from the $\eta_c \to \kskpi$ and $\etac \to \Kp \Km \piz$ MIPWA. Phases are determined relative to the $(K\pi \ \mathcal{S}$-wave) $\Kbar$ amplitude which is fixed to $\pi/2$ at 1.45 \gevcc.}
\label{tab:tab1}
\begin{center}
  \begin{tabular}{|l | r@{}c@{}r | r@{}c@{}r | r@{}c@{}r | r@{}c@{}r|}
    \hline \\ [-2.3ex]
      &   \multicolumn{6}{c|} {$\etac \to \kskpi$} &   \multicolumn{6}{c|}{ \etacpiz  }  \cr
   \hline \\ [-2.3ex]
  Amplitude & \multicolumn{3}{c|} {Fraction (\%)}  & \multicolumn{3}{c|}{Phase (rad)} & \multicolumn{3}{c|}{Fraction (\%)} & \multicolumn{3}{c|}{Phase (rad)}\cr
   \hline \\ [-2.3ex]
$(K\pi \ \mathcal{S}$-wave) $\Kbar$ & 107.3 $\pm$  & \, 2.6 $\pm$  & \, 17.9 & &  fixed  &  & 125.5 $\pm$  & \, 2.4 $\pm$  & \, 4.2 & & fixed  &  \cr
$a_0(980) \pi$ & 0.8 $\pm$  & \, 0.5 $\pm$  & \, 0.8 & 1.08 $\pm$  & \, 0.18  $\pm$  & \, 0.18 & 0.0 $\pm$  & \, 0.1 $\pm$  & \, 1.7 &  & - & \cr
$a_0(1450) \pi$ & 0.7 $\pm$  & \, 0.2 $\pm$  & \, 1.4 & 2.63 $\pm$  & \, 0.13 $\pm$  & \, 0.17 & 1.2 $\pm$  & \, 0.4 $\pm$  & \, 0.7 & 2.90 $\pm$  & \, 0.12 $\pm$  & \, 0.25\cr
$a_0(1950) \pi$ & 3.1 $\pm$  & \, 0.4 $\pm$  & \, 1.2 & $-$1.04 $\pm$  & \, 0.08 $\pm$  & \, 0.77& 4.4 $\pm$  & \, 0.8 $\pm$  & \, 0.8& $-$1.45 $\pm$  & \, 0.08 $\pm$  & \, 0.27\cr
$a_2(1320) \pi$& 0.2 $\pm$  & \, 0.1 $\pm$  & \, 0.1 & 1.85 $\pm$  & \, 0.20 $\pm$  & \, 0.20 & 0.6 $\pm$  & \, 0.2 $\pm$  & \, 0.3& 1.75 $\pm$  & \, 0.23 $\pm$  & \, 0.42\cr
$K^*_2(1430) \Kbar$ & 4.7 $\pm$  & \, 0.9 $\pm$  & \, 1.4 & 4.92 $\pm$  & \, 0.05 $\pm$  & \, 0.10 & 3.0 $\pm$  & \, 0.8 $\pm$  & \, 4.4 & 5.07 $\pm$  & \, 0.09 $\pm$  & \, 0.30\cr
 \hline \\ [-2.3ex]
Total & 116.8 $\pm$  & \, 2.8  $\pm$ & \, 18.1 &   & &    & 134.8 $\pm$  & \, 2.7 $\pm$ & \, 6.4 & & &   \cr
$-$ $2\log {\cal L}$ & \multicolumn{3}{c|} {$-$4314.2} & & & & \multicolumn{3}{c|}{$-$2339} & & & \cr
$\chi^2/N_{\rm cells}$ & \multicolumn{3}{c|} {301/254=1.17} & & & &  \multicolumn{3}{c|}{283.2/233=1.22} & & & \cr
\hline
\end{tabular}
\end{center}
\end{table*}

The table lists also a significant contribution from the $a_0(1950) \pi$ amplitude, where $a_0(1950)^+ \to \KS \Kp$ is a new
resonance. 
We also test the spin-2 hypothesis for this contribution by replacing the amplitude for $a_0 \to K^0_S K^+$ with an $a_2 \to K^0_S K^+$ amplitude with parameter values left free in the fit.
In this case no physical solution is found inside the allowed ranges of the parameters, and the additional contribution is found consistent with zero. This new state has isospin one, and the spin-0 assignment is preferred over that of spin-2.

\begin{figure*}
\begin{center}
\includegraphics[width=16cm]{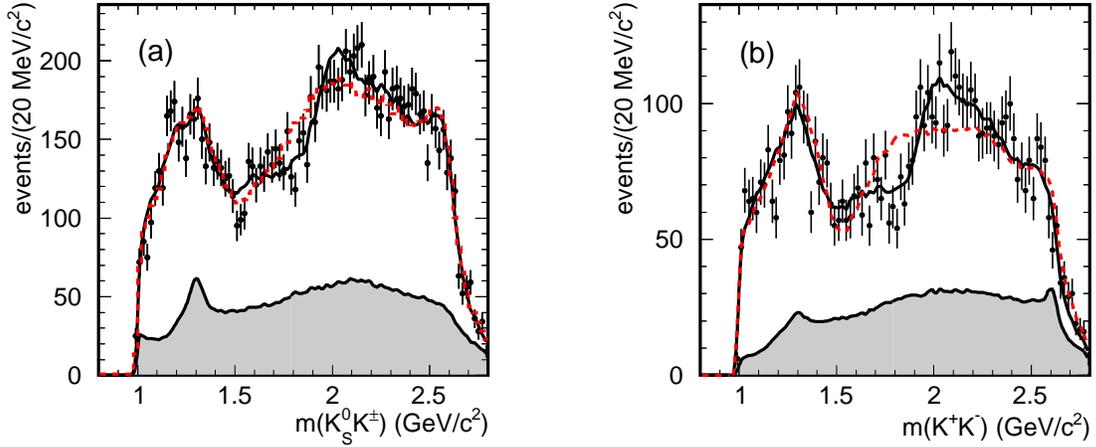}
\caption{The mass projections (a) $\KS \Kpm$ from $\etac \to \kskpi$ and (b) $\Kp \Km$ from $\etac \to \kkpiz$. The histograms show the MIPWA fit projections with (solid, black) and without (dashed, red) the presence of the $a_0(1950)^+ \to \KS \Kpm$ resonance.  The shaded regions show the background estimates obtained by interpolating the results of the Dalitz plot analyses of the sideband regions.}
\label{fig:fig7}
\end{center}
\end{figure*}

A fit without this state gives a poor description of the high mass $\KS \Kp$ projection, as can be seen in Fig.~\ref{fig:fig7}(a). We obtain $-2\log {\cal L} = -$4252.9 and
$\chi^2/N_{\rm cells}=1.33$ for this fit.
We then include in the MIPWA a new scalar resonance decaying to $\KS \Kp$ with free parameters. 
We obtain 
$\Delta (\log {\cal L})=61$ and  $\Delta \chi^2=38$ for an increase of four new parameters.
We estimate the significance for the $a_0(1950)$ resonance using the fitted 
fraction divided by its statistical and systematic errors added in quadrature, and obtain $2.5\sigma$.
Since interference effects may also contribute to the significance, this procedure gives a conservative estimate.
The systematic uncertainties associated with the $a_0(1950)$ state are described below.
The fitted parameter values for this state are given in Table~\ref{tab:tab2}.
We note that we obtain $\chi^2/N_{\rm cells}=1.17$ for this final fit, indicating good description of the data.
The fit projections on the three squared masses from the MIPWA are shown in Fig.~\ref{fig:fig5}, and they indicate that the description of the data is quite good.

\begin{table}
\caption{Fitted $a_0(1950)$ parameter values for the two \etac decay modes.}
\label{tab:tab2}
\begin{center}
\begin{tabular}{|l|c|r@{}c@{}r|}
  \hline \\ [-2.3ex]
Final state & Mass (\mevcc) & \multicolumn{3}{c|}{Width (\mev)}\cr
\hline \\ [-2.3ex]
$\etac \to \kskpi$ & 1949 $\pm$ 32 $\pm$ 76 & 265 $\pm$ & \, 36 $\pm$ & \,110\cr
\etacpiz & 1927 $\pm$ 15 $\pm$ 23 & 274 $\pm$ & \, 28 $\pm$ & \, 30 \cr
\hline \\ [-2.3ex]
Weighted mean & 1931 $\pm$ 14 $\pm$ 22 & 271 $\pm$ & \, 22 $\pm$ & \, 29\cr
\hline
\end{tabular}
\end{center}
\end{table}

\begin{figure*}
\begin{center}
\includegraphics[width=16cm]{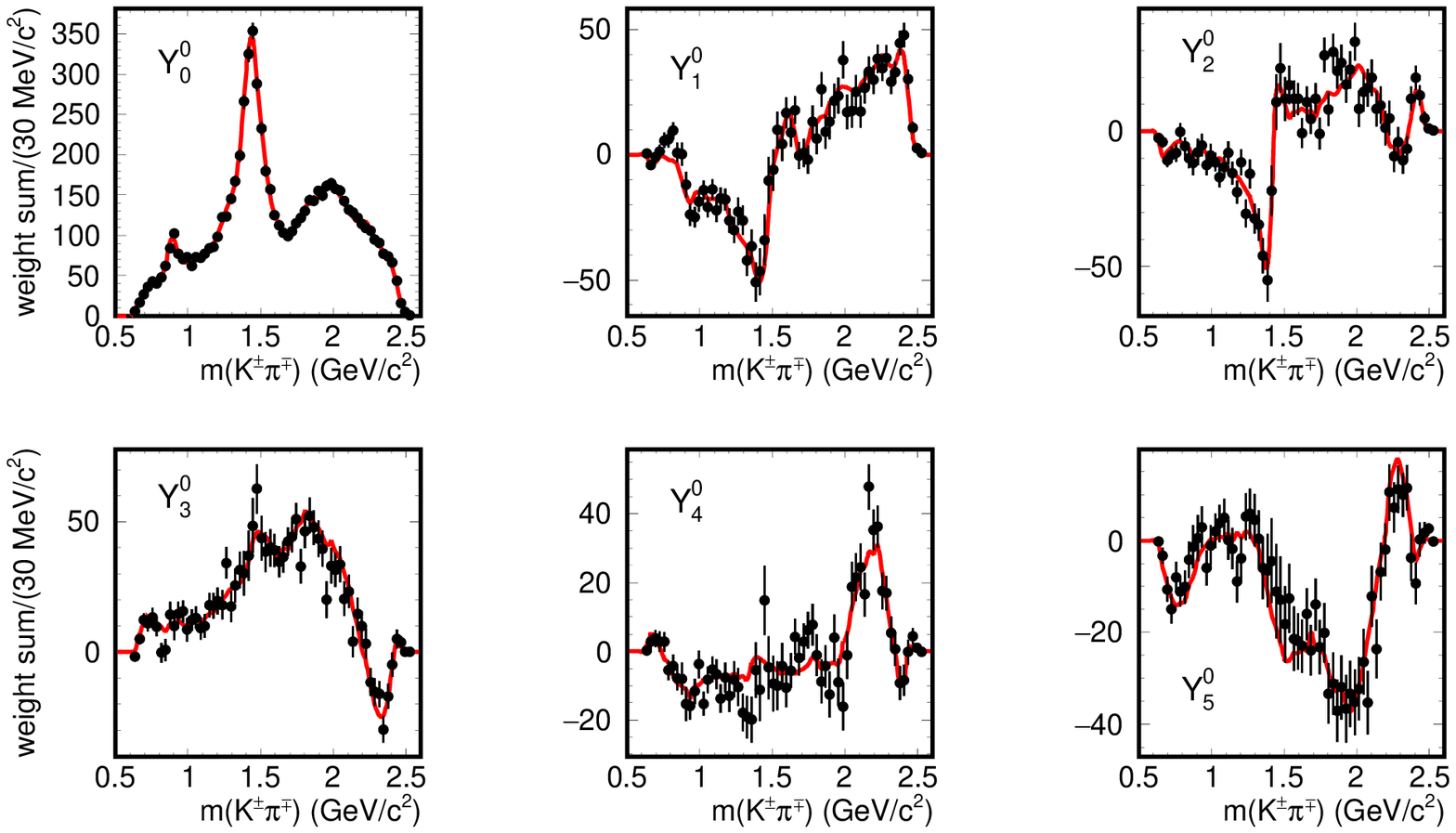}
\caption{Legendre polynomial moments for $\etac \to \kskpi$ as functions of $K \pi$ mass, and combined for $\Kpm \pimp$ and  $\KS \pimp$; the superimposed curves result from the Dalitz plot fit described in the text.}
\label{fig:fig8}
\end{center}
\end{figure*}

\begin{figure*}
\begin{center}
\includegraphics[width=16cm]{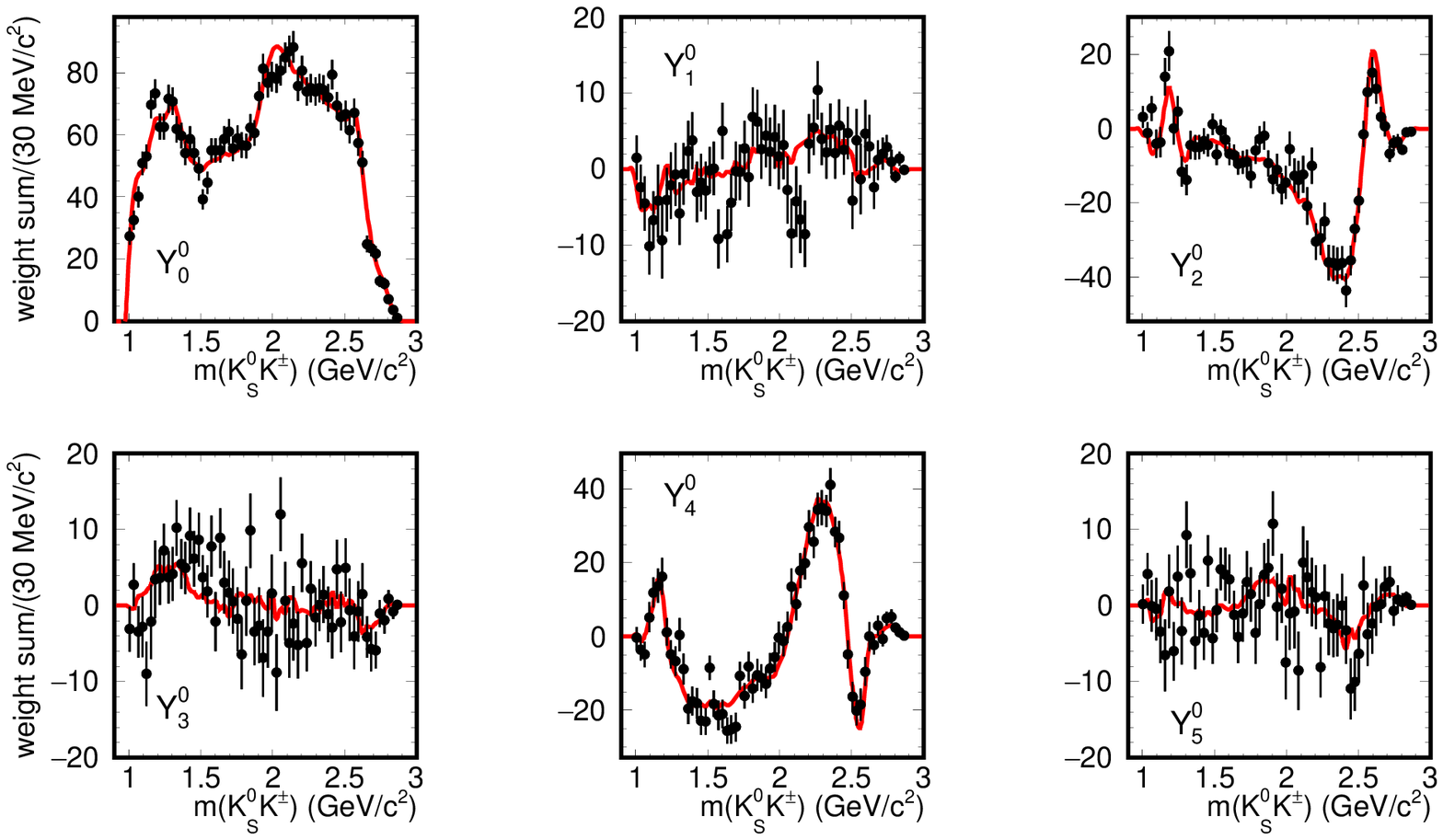}
\caption{Legendre polynomial moments for $\etac \to \kskpi$ as a function of $\KS \Kpm$ mass, the superimposed curves result from the Dalitz plot fit described in the text.}
\label{fig:fig9}
\end{center}
\end{figure*}

We compute the uncorrected Legendre polynomial moments $\langle Y^0_L \rangle$ in each $\Kp \pim$, $\KS \pim$ and $\KS \Kp$  mass interval by weighting each event by the relevant $Y^0_L(\cos \theta)$ function.
These distributions are shown in Fig.~\ref{fig:fig8} as functions of $K \pi$ mass after combining $\Kp \pim$ and  $\KS \pim$, and in Fig.~\ref{fig:fig9} as functions of $\KS \Kp$ mass. We also compute the expected Legendre polynomial moments from the weighted MC events and compare with the experimental distributions. We observe good agreement for all the distributions, which indicates that the fit is able to reproduce the local structures apparent in the Dalitz plot.

We compute the following systematic uncertainties on the $I=1/2$  $K \pi$ $\mathcal{S}$-wave amplitude and phase. The different contributions are added in quadrature.
\begin{itemize}
\item{} Starting from the solution found by the fit, we generate MC simulated events which are fitted using a MIPWA. In this way we estimate the bias introduced by the fitting method.
\item{} The fit is performed by interpolating the  $K \pi$ $\mathcal{S}$-wave amplitude and phase using a cubic spline.
\item{} We remove low-significance contributions, such as those from the $a_0(980)$ and $a_2(1320)$ resonances.
\item{} We vary the signal purity up and down according to its statistical uncertainty.
\item{} The effect of the efficiency variation as a function of $K \Kbar \pi$ mass is evaluated by computing separate efficiencies
  in the regions below and above the $\eta_c$ mass.
\end{itemize}

These additional fits also allow the computation of systematic uncertainties on the amplitude fraction and phase values, as well as on the parameter values for the $a_0(1950)$ resonance; these are summarized in Table~\ref{tab:a0_sys}. In the evaluation of overall systematic uncertainties, all effects are assumed to be uncorrelated, and are added in quadrature.

The measured amplitude and phase values of the $I=1/2$  $K \pi$ $\mathcal{S}$-wave as functions of mass obtained from the MIPWA of $\etac \to \kskpi$ are shown in Table~\ref{tab:tab6}. Interval 14 of the $K \pi$ mass contains the fixed amplitude and phase values.

\begin{table*}
  \caption{Systematic uncertainties on the $a_0(1950)$ parameter values from the two \etac decay modes.}
  \label{tab:a0_sys}
       \begin{center}
         \begin{tabular}{|l|r|r|r|r|r|r|}
           \hline \\ [-2.3ex]
        &  \multicolumn{3}{c|}{$\etac \to \kskpi$} & \multicolumn{3}{c|}{$\etac \to \Kp \Km \piz$} \cr
      \hline \\ [-2.3ex]   
    Effect & Mass  & Width  & Fraction (\%) & Mass  & Width & Fraction (\%)\cr
      &  (\mevcc) &  (MeV) &  & (\mevcc) &  (MeV) & \cr   
        \hline \\ [-2.3ex]
Fit bias & 11 & 22 & 0.5 & 1 & 10 & 0.5 \cr
Cubic spline & 24 & 79 & 0.6 & 14 & 9 & 0.2\cr
Marginal components  & 70 & 72 & 0.0 & 2 & 8 & 0.3 \cr
\etac purity  & 3 & 16  & 1.0 & 18 & 26 & 0.4 \cr
Efficiency & 11 & 8  & 0.2 & 1  & 15  & 0.2 \cr
 \hline \\ [-2.3ex]
Total & 76 & 110 & 1.3 & 23 & 30 & 0.8\cr
\hline
  \end{tabular}
 \end{center} 
\end{table*}

\begin{table*}
  \caption{Measured amplitude and phase values for the $I=1/2$  $K \pi$ $\mathcal{S}$-wave as functions of mass obtained from the MIPWA of $\etac \to \kskpi$ and $\eta_c \to \Kp \Km \piz$. The first error is statistical, the second systematic. The amplitudes and phases in the mass interval 14
  are fixed to constant values.}
\label{tab:tab6}
\begin{center}
\begin{tabular}{|c|c|c|c|c|c|}
 \hline \\ [-2.3ex]
  \multicolumn{2}{|c}{} & \multicolumn{2}{|c}{$\etac \to \kskpi$} & \multicolumn{2}{|c|}{$\etac \to \Kp \Km \piz$} \cr
 \hline \\ [-2.3ex]
N & $K \pi$ mass & Amplitude & Phase (rad) & Amplitude & Phase (rad)\cr
 \hline \\ [-2.3ex]
   1 & 0.67 & $ \ 0.119 \pm  0.100 \ \pm 0.215$ & $ \ \ 0.259 \pm  0.577 \ \pm 1.290$ & $ \ 0.154 \pm  0.350 \ \pm 0.337$ & $ \ \ 3.786 \pm  1.199 \ \pm 0.857$ \cr
   2 & 0.73 & $ \ 0.103 \pm  0.043 \ \pm 0.113$ & $  -0.969 \pm  0.757 \ \pm 1.600$ & $ \ 0.198 \pm  0.124 \ \pm 0.216$ & $ \ \ 3.944 \pm  0.321 \ \pm 0.448$ \cr
   3 & 0.79 & $ \ 0.158 \pm  0.086 \ \pm 0.180$ & $ \ \  0.363 \pm  0.381 \ \pm 1.500$ & $ \ 0.161 \pm  0.116 \ \pm 0.098$ & $ \ \ 1.634 \pm  0.584 \ \pm 0.448$ \cr
   4 & 0.85 & $ \ 0.232 \pm  0.128 \ \pm 0.214$ & $ \ \  0.448 \pm  0.266 \ \pm 1.500$ & $ \ 0.125 \pm  0.118 \ \pm 0.031$ & $ \ \ 3.094 \pm  0.725 \ \pm 0.448$ \cr
   5 & 0.91 & $ \ 0.468 \pm  0.075 \ \pm 0.194$ & $ \ \  0.091 \pm  0.191 \ \pm 0.237$ & $ \ 0.307 \pm  0.213 \ \pm 0.162$ & $ \ \ 0.735 \pm  0.326 \ \pm 0.255$ \cr
   6 & 0.97 & $ \ 0.371 \pm  0.083 \ \pm 0.129$ & $ \ \  0.276 \pm  0.156 \ \pm 0.190$ & $ \ 0.528 \pm  0.121 \ \pm 0.055$ & $  -0.083 \pm  0.178 \ \pm 0.303$ \cr
   7 & 1.03 & $ \ 0.329 \pm  0.071 \ \pm 0.102$ & $ \ \  0.345 \pm  0.164 \ \pm 0.273$ & $ \ 0.215 \pm  0.191 \ \pm 0.053$ & $ \ \ 0.541 \pm  0.320 \ \pm 0.638$ \cr
   8 & 1.09 & $ \ 0.343 \pm  0.062 \ \pm 0.062$ & $ \ \  0.449 \pm  0.196 \ \pm 0.213$ & $ \ 0.390 \pm  0.146 \ \pm 0.046$ & $ \ \ 0.254 \pm  0.167 \ \pm 0.144$ \cr
   9 & 1.15 & $ \ 0.330 \pm  0.070 \ \pm 0.081$ & $ \ \  0.687 \pm  0.167 \ \pm 0.221$ & $ \ 0.490 \pm  0.135 \ \pm 0.089$ & $ \ \ 0.618 \pm  0.155 \ \pm 0.099$ \cr
  10 & 1.21 & $ \ 0.450 \pm  0.059 \ \pm 0.042$ & $ \ \  0.696 \pm  0.156 \ \pm 0.226$ & $ \ 0.422 \pm  0.092 \ \pm 0.102$ & $ \ \ 0.723 \pm  0.242 \ \pm 0.267$ \cr
  11 & 1.27 & $ \ 0.578 \pm  0.048 \ \pm 0.112$ & $ \ \  0.785 \pm  0.208 \ \pm 0.358$ & $ \ 0.581 \pm  0.113 \ \pm 0.084$ & $ \ \ 0.605 \pm  0.186 \ \pm 0.166$ \cr
  12 & 1.33 & $ \ 0.627 \pm  0.047 \ \pm 0.053$ & $ \ \  0.986 \pm  0.153 \ \pm 0.166$ & $ \ 0.643 \pm  0.106 \ \pm 0.039$ & $ \ \ 1.330 \pm  0.264 \ \pm 0.130$ \cr
  13 & 1.39 & $ \ 0.826 \pm  0.047 \ \pm 0.105$ & $ \ \  1.334 \pm  0.155 \ \pm 0.288$ & $ \ 0.920 \pm  0.153 \ \pm 0.056$ & $ \ \ 1.528 \pm  0.161 \ \pm 0.160$ \cr
  \textcolor{red}{14} & \textcolor{red}{1.45} & \textcolor{red}{$ \ 1.000 $} & \textcolor{red}{$ \ 1.570  $} & \textcolor{red}{$ \ 1.000   $} & \textcolor{red}{$ \ 1.570  $} \cr
  15 & 1.51 & $ \ 0.736 \pm  0.031 \ \pm 0.059$ & $ \ \ 1.918 \pm  0.153 \ \pm 0.132$ & $ \ 0.750 \pm  0.118 \ \pm 0.076$ & $ \ \ 1.844 \pm  0.149 \ \pm 0.048$ \cr
  16 & 1.57 & $ \ 0.451 \pm  0.025 \ \pm 0.053$ & $ \ \ 2.098 \pm  0.202 \ \pm 0.277$ & $ \ 0.585 \pm  0.099 \ \pm 0.047$ & $ \ \ 2.128 \pm  0.182 \ \pm 0.110$ \cr
  17 & 1.63 & $ \ 0.289 \pm  0.029 \ \pm 0.065$ & $ \ \ 2.539 \pm  0.292 \ \pm 0.180$ & $ \ 0.366 \pm  0.079 \ \pm 0.052$ & $ \ \ 2.389 \pm  0.230 \ \pm 0.213$ \cr
  18 & 1.69 & $ \ 0.159 \pm  0.036 \ \pm 0.089$ & $ \ \ 1.566 \pm  0.308 \ \pm 0.619$ & $ \ 0.312 \pm  0.074 \ \pm 0.043$ & $ \ \ 1.962 \pm  0.195 \ \pm 0.150$ \cr
  19 & 1.75 & $ \ 0.240 \pm  0.034 \ \pm 0.067$ & $ \ \ 1.962 \pm  0.331 \ \pm 0.655$ & $ \ 0.427 \pm  0.093 \ \pm 0.063$ & $ \ \ 1.939 \pm  0.150 \ \pm 0.182$ \cr
  20 & 1.81 & $ \ 0.381 \pm  0.031 \ \pm 0.059$ & $ \ \ 2.170 \pm  0.297 \ \pm 0.251$ & $ \ 0.511 \pm  0.094 \ \pm 0.063$ & $ \ \ 2.426 \pm  0.156 \ \pm 0.277$ \cr
  21 & 1.87 & $ \ 0.457 \pm  0.035 \ \pm 0.085$ & $ \ \ 2.258 \pm  0.251 \ \pm 0.284$ & $ \ 0.588 \pm  0.098 \ \pm 0.080$ & $ \ \ 2.242 \pm  0.084 \ \pm 0.210$ \cr
  22 & 1.93 & $ \ 0.565 \pm  0.042 \ \pm 0.067$ & $ \ \ 2.386 \pm  0.255 \ \pm 0.207$ & $ \ 0.729 \pm  0.114 \ \pm 0.095$ & $ \ \ 2.427 \pm  0.098 \ \pm 0.254$ \cr
  23 & 1.99 & $ \ 0.640 \pm  0.044 \ \pm 0.055$ & $ \ \ 2.361 \pm  0.228 \ \pm 0.092$ & $ \ 0.777 \pm  0.119 \ \pm 0.075$ & $ \ \ 2.306 \pm  0.102 \ \pm 0.325$ \cr
  24 & 2.05 & $ \ 0.593 \pm  0.046 \ \pm 0.065$ & $ \ \ 2.329 \pm  0.235 \ \pm 0.268$ & $ \ 0.775 \pm  0.134 \ \pm 0.075$ & $ \ \ 2.347 \pm  0.107 \ \pm 0.299$ \cr
  25 & 2.11 & $ \ 0.614 \pm  0.057 \ \pm 0.083$ & $ \ \ 2.421 \pm  0.230 \ \pm 0.169$ & $ \ 0.830 \pm  0.134 \ \pm 0.078$ & $ \ \ 2.374 \pm  0.105 \ \pm 0.199$ \cr
  26 & 2.17 & $ \ 0.677 \pm  0.067 \ \pm 0.117$ & $ \ \ 2.563 \pm  0.218 \ \pm 0.137$ & $ \ 0.825 \pm  0.140 \ \pm 0.070$ & $ \ \ 2.401 \pm  0.127 \ \pm 0.189$ \cr
  27 & 2.23 & $ \ 0.788 \pm  0.085 \ \pm 0.104$ & $ \ \ 2.539 \pm  0.228 \ \pm 0.241$ & $ \ 0.860 \pm  0.158 \ \pm 0.123$ & $ \ \ 2.296 \pm  0.131 \ \pm 0.297$ \cr
  28 & 2.29 & $ \ 0.753 \pm  0.097 \ \pm 0.125$ & $ \ \ 2.550 \pm  0.234 \ \pm 0.168$ & $ \ 0.891 \pm  0.167 \ \pm 0.133$ & $ \ \ 2.320 \pm  0.131 \ \pm 0.273$ \cr
  29 & 2.35 & $ \ 0.646 \pm  0.096 \ \pm 0.118$ & $ \ \ 2.315 \pm  0.241 \ \pm 0.321$ & $ \ 0.994 \pm  0.202 \ \pm 0.076$ & $ \ \ 2.297 \pm  0.153 \ \pm 0.197$ \cr
  30 & 2.41 & $ \ 0.789 \pm  0.184 \ \pm 0.187$ & $ \ \ 2.364 \pm  0.336 \ \pm 0.199$ & $ \ 0.892 \pm  0.322 \ \pm 0.098$ & $ \ \ 2.143 \pm  0.292 \ \pm 0.393$ \cr
\hline
\end{tabular}
\end{center}
\end{table*}
 
\subsection{Dalitz plot analysis of {\boldmath$\protect \etactokkpi$} using an isobar model}

We perform a Dalitz plot analysis of \etactokkpi using a standard isobar model, where all resonances
are modeled as BW functions multiplied by the corresponding angular functions. In this case the $K \pi$ $\mathcal{S}$-wave is represented by a superposition of interfering $K^*_0(1430)$, \Kstarzp, non-resonant (NR), and possibly $\kappa(800)$ contributions.
The NR contribution is parametrized as an amplitude that is constant in magnitude and phase
over the Dalitz plot.
In this fit the $K^*_0(1430)$ parameters
are taken from Ref.~\cite{etakk}, while all other parameters are fixed to PDG values. We also add the $a_0(1950)$
resonance with parameters obtained from the MIPWA analysis.

For the description of the \etac signal, amplitudes are added one by one to ascertain the associated increase of the likelihood value and decrease of the \mbox{2-D} $\chi^2$. 
Table~\ref{tab:tab4} summarizes the fit results for the amplitude fractions and phases.
The high value of $\chi^2/N_{\rm cells}=1.82$ (to be compared with $\chi^2/N_{\rm cells}=1.17$) indicates a poorer description of the data than that obtained with the MIPWA method. Including the $\kappa(800)$ resonance does not improve the fit quality. If included, it gives a fit fraction of $(0.8 \pm 0.5)$\%.

The Dalitz plot analysis shows a dominance of scalar meson amplitudes, with small contributions from spin-two resonances. The $K^*(892)$ contribution is consistent with originating entirely from background. Other spin-1 $K^*$ resonances have been included in the fit, 
but their contributions have been found to be
consistent with zero. We note the presence of a sizeable non-resonant contribution. However, in this case the sum of the fractions is significantly lower than 100\%, indicating important interference effects.
Fitting the data without the NR contribution gives a much poorer description, with $-2\log {\cal L}=-$4115 and $\chi^2/N_{\rm cells}=2.32$.

We conclude that the $\etac \to \kskpi$ Dalitz plot is not well-described by an isobar model in which the $K \pi$ $\mathcal{S}$-wave is modeled as a superposition of Breit-Wigner functions. A more complex approach is needed, and the MIPWA is able to describe
this amplitude without the need for a specific model.

\begin{table}
\caption{Results from the \etactokkpi Dalitz plot analysis using an isobar model. The listed uncertainties are statistical only.}
\label{tab:tab4}
\begin{center}
  \begin{tabular}{|l|r@{}c|r@{}c|}
     \hline \\ [-2.3ex]
 Amplitude & \multicolumn{2}{c|}{Fraction \%} & \multicolumn{2}{c|}{Phase (rad)}\cr
  \hline \\ [-2.3ex]
$K^*_0(1430) \Kbar$ & 40.8 $\pm$ & \, 2.2 &  0. & \cr
$K^*_0(1950) \Kbar$ & 14.8 $\pm$ & \, 1.7 & $-$1.00 $\pm$ & \, 0.07 \cr
NR & 18.0 $\pm$ & \, 2.5 & 1.94 $\pm$ & \, 0.09 \cr
$a_0(980) \pi$ & 10.5 $\pm$ & \, 1.2 & 0.94 $\pm$ & \, 0.12 \cr
$a_0(1450) \pi$ & 1.7 $\pm$ & \, 0.5 & 2.94 $\pm$ & \, 0.13 \cr
$a_0(1950) \pi$ & 0.7 $\pm$ & \, 0.2 & $-$1.76 $\pm$ & \, 0.24 \cr
$a_2(1320) \pi$ & 0.2 $\pm$ & \, 0.2 & $-$0.53 $\pm$ & \, 0.42 \cr
$K^*_2(1430) \Kbar$ & 2.3 $\pm$ & \, 0.7 & $-$1.55 $\pm$ & \, 0.11 \cr
 \hline \\ [-2.3ex]
Total &  88.8 $\pm$ & \, 4.3 & & \cr
$-2\log {\cal L}$ &  $-$4290.7 & \,  & &  \cr
$\chi^2/N_{\rm cells}$ & 467/256=1.82 & \,   & & \cr
\hline
\end{tabular}
\end{center}
\end{table}

\section{Dalitz plot analysis of {\boldmath$\protect \etacpiz$}\ }

The \etacpiz Dalitz plot~\cite{etakk} is very similar to that for $\etac \to \kskpi$ decays. It is dominated by uniformly populated  bands at
the $K^*_0(1430)$ resonance position in $\Kp \piz$ and $\Km \piz$ mass squared. It also shows a broad diagonal structure indicating
the presence of $a_0$ or $a_2$ resonance contributions. The Dalitz plot projections are shown in Fig.~\ref{fig:fig10}.

The \etacpiz Dalitz plot analysis using the isobar model has been performed already in Ref.~\cite{etakk} . It was found that
the model does not give a perfect description of the data. In this section we  obtain a new measurement
of the $K \pi$ $\mathcal{S}$-wave by making use of the MIPWA method. In this way we also perform a cross-check of the results
obtained from the \etactokkpi analysis, since analyses of the two $\etac$ decay modes should give consistent results, given
the absence of I=3/2 $K \pi$ amplitude contributions.

\subsection{MIPWA of {\boldmath$\protect \etacpiz$}}

We perform a MIPWA of \etacpiz decays using the same model and the same mass grid as for \etactokkpi.
As for the previous case we obtain a better description of the data if we include an additional $a_0(1950)$ resonance, whose parameter values are listed in Table~\ref{tab:tab2}. We observe good agreement between the parameter values obtained from the two \etac decay modes.
The table also lists parameter values obtained as the weighted mean of the two measurements.
Table~\ref{tab:tab1} gives the fitted fractions from the MIPWA fit.

\begin{figure*}
\begin{center}
\includegraphics[width=18cm]{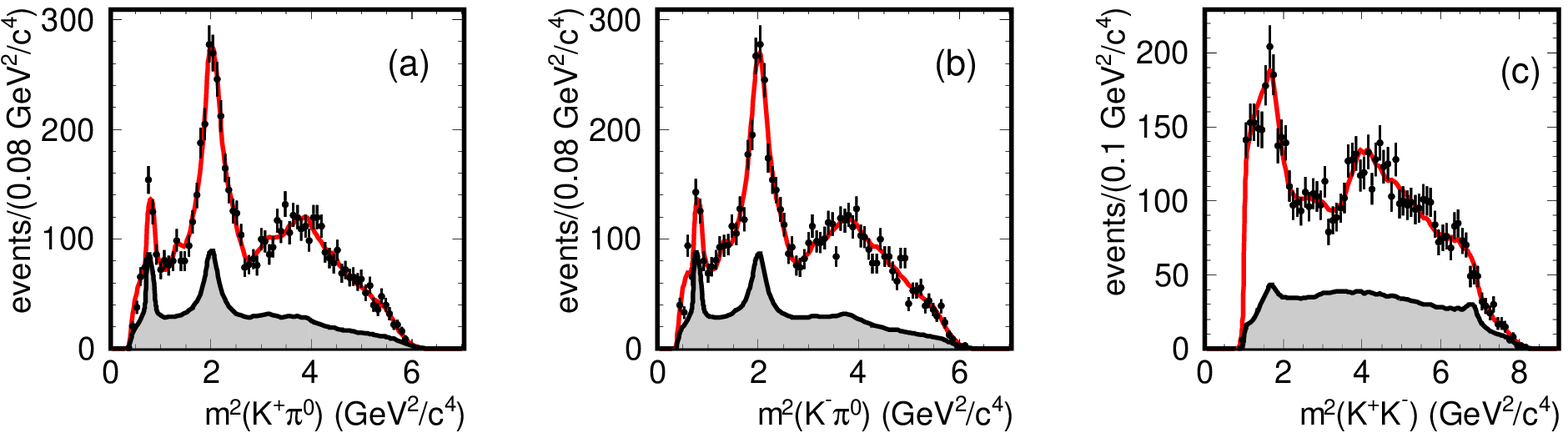}
\caption{The $\etac \to \Kp \Km \piz$ Dalitz plot projections, (a) $m^2(\Kp \piz)$, (b) $m^2(\Km \piz)$, and (c) $m^2(\Kp \Km)$. The superimposed curves result from the MIPWA described in the text. The shaded regions show the
background estimates obtained  by interpolating the results of the Dalitz plot analyses of the sideband regions.}
\label{fig:fig10}
\end{center}
\end{figure*}

We obtain a good description of the data, as evidenced by the value $\chi^2/N_{\rm cells}=1.22$, and observe
the $a_0(1950)$ state with a significance of $4.2\sigma$.
The fit projections on the $\Kp \piz$, $\Km \piz$, and $\Kp \Km$ squared mass distributions are shown in Fig.~\ref{fig:fig10}. As previously, there is
a dominance of the ($K \pi$ $\mathcal{S}$-wave) $\Kbar$ amplitude, with a significant $K^*_2(1430) \Kbar$ amplitude, and small contributions from $a_0 \pi$ amplitudes. We observe good agreement between fractions and relative phases of the
amplitudes between the \etactokkpi and \etacpiz decay modes.
Systematic uncertainties are evaluated as discussed in Sec. VI.A.

\begin{figure*}
\begin{center}
\includegraphics[width=16cm]{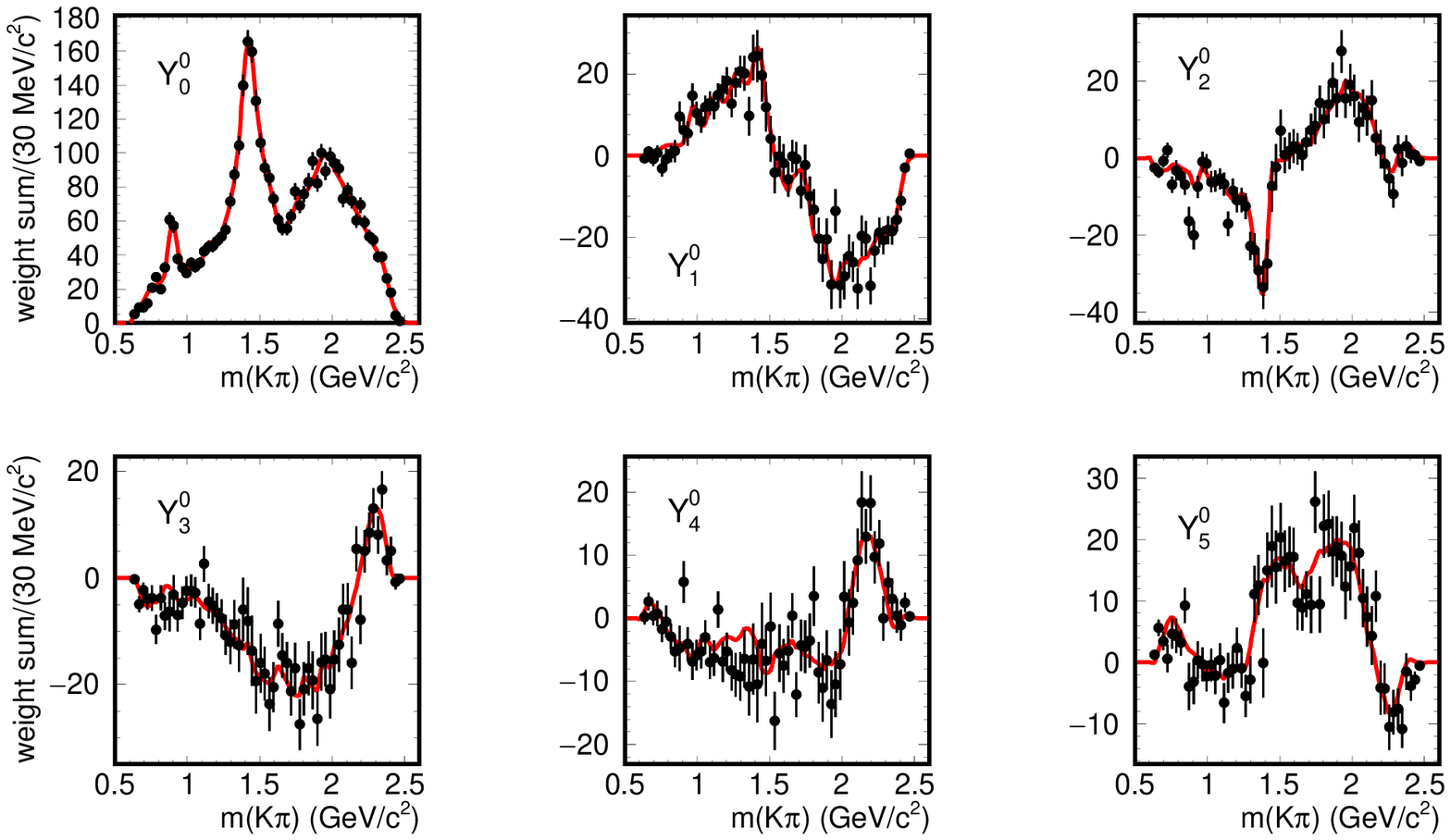}
\caption{Legendre polynomial moments for $\etac \to \Kp \Km \piz$ as functions of $K \pi$ mass, combined for $\Kp \piz$ and  $\Km \piz$. The superimposed curves result from the Dalitz plot fit described in the text.}
\label{fig:fig11}
\end{center}
\end{figure*}

\begin{figure*}
\begin{center}
\includegraphics[width=16cm]{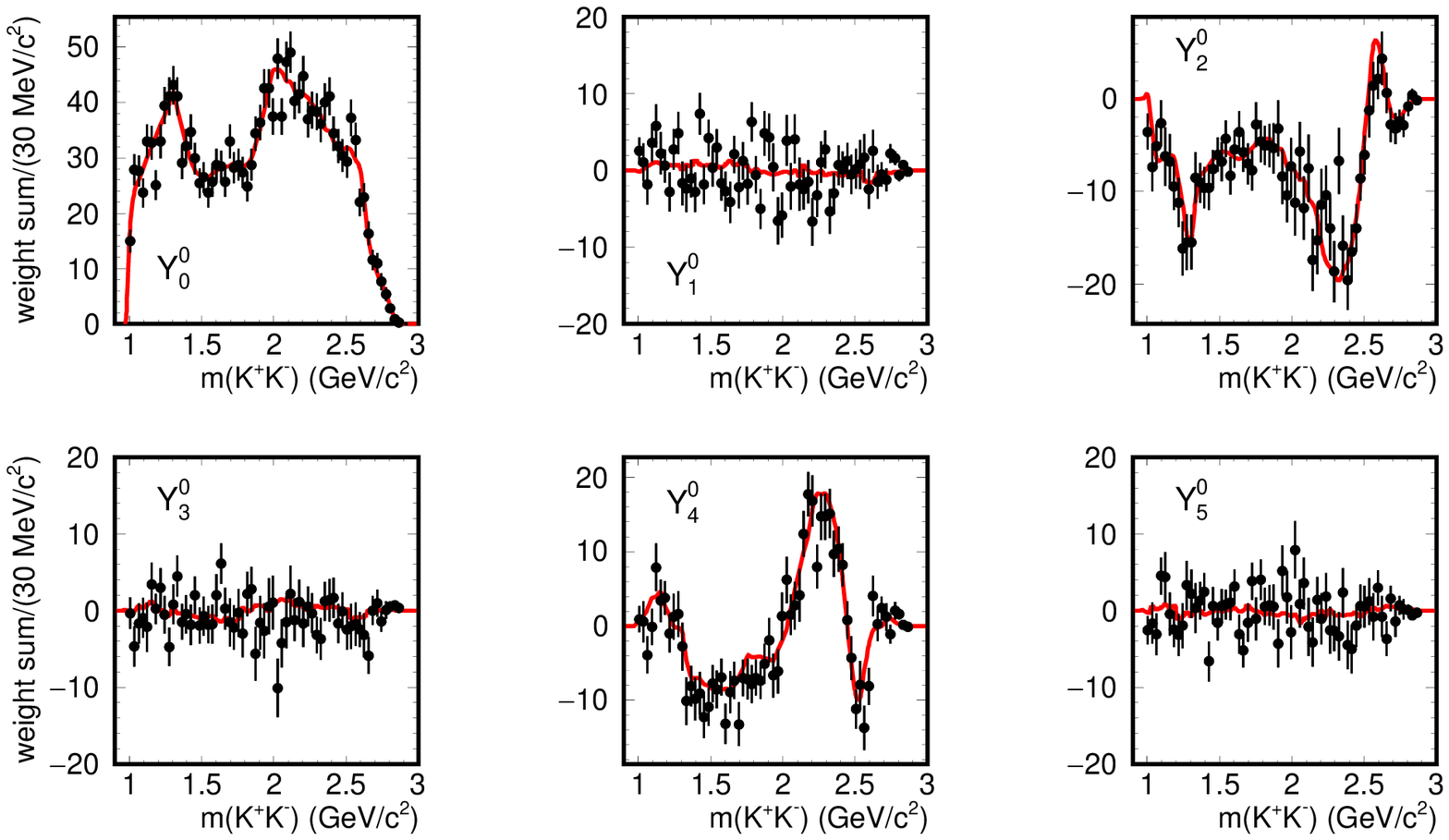}
\caption{Legendre polynomial moments for $\etac \to \Kp \Km \piz$ as a function of $\Kp \Km$ mass. The superimposed curves result from the Dalitz plot fit described in the text.}
\label{fig:fig12}
\end{center}
\end{figure*}

We compute the uncorrected Legendre polynomial moments $\langle Y^0_L \rangle$ in each $\Kp \piz$, $\Km \piz$ and $\Kp \Km$  mass interval by weighting each event by the relevant $Y^0_L(\cos \theta)$ function.
These distributions are shown in Fig.~\ref{fig:fig11} as functions of  $K \pi$ mass, combined for $\Kp \piz$ and  $\Km \piz$, and in Fig.~\ref{fig:fig12} as functions of $\Kp \Km$ mass. We also compute the expected Legendre polynomial moments from the weighted MC events and compare with the experimental distributions. We observe good agreement for all the distributions, which indicates that also in this case the fit is able to reproduce the local structures apparent in the Dalitz plot.

\begin{figure*}
\begin{center}
\includegraphics[width=16cm]{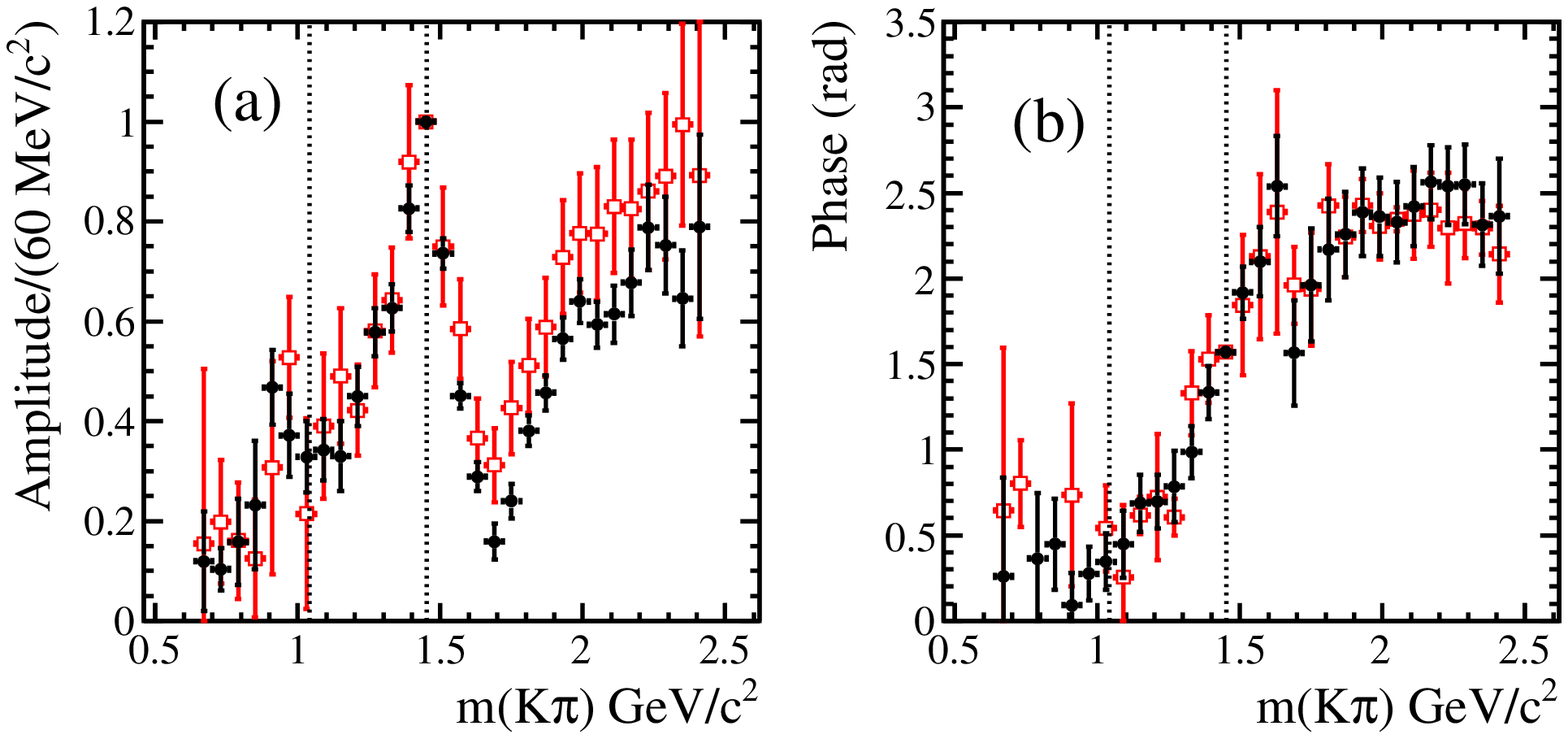}
\caption{The $I=1/2$  $K \pi$ $\mathcal{S}$-wave amplitude (a) and phase (b) from $\etac \to \kskpi$ (solid (black) points) and \etacpiz
  (open (red) points); only statistical uncertainties are shown. The dotted lines indicate the $K \eta$ and $K \eta'$ thresholds.}
\label{fig:fig13}
\end{center}
\end{figure*}

\section{The $I=1/2$  {\boldmath$\protect  K \pi \ \mathcal{S}$}-wave amplitude and phase}

Figure~\ref{fig:fig13} displays the measured $I=1/2$  $K \pi$ $\mathcal{S}$-wave amplitude and phase from both $\etac \to \kskpi$ and \etacpiz.
We observe good agreement between the amplitude and phase values obtained from the two measurements.

The main features of the amplitude (Fig.~\ref{fig:fig13}(a)) can be explained by the presence of a clear peak related to the $K^*_0(1430)$ resonance
which shows a rapid drop around 1.7 \gevcc, where a broad structure is present which can be related to the $K^*_0(1950)$ resonance.
There is some indication of feedthrough from the $K^*(892)$ background.
The phase motion (Fig.~\ref{fig:fig13}(b)) shows the expected behavior for the resonance phase, which varies by about $\pi$ in the $K^*_0(1430)$ resonance region. The phase shows a drop around 1.7 \gevcc\ related to interference with the $K^*_0(1950)$ resonance.

\begin{figure*}
\begin{center}
\includegraphics[width=18cm]{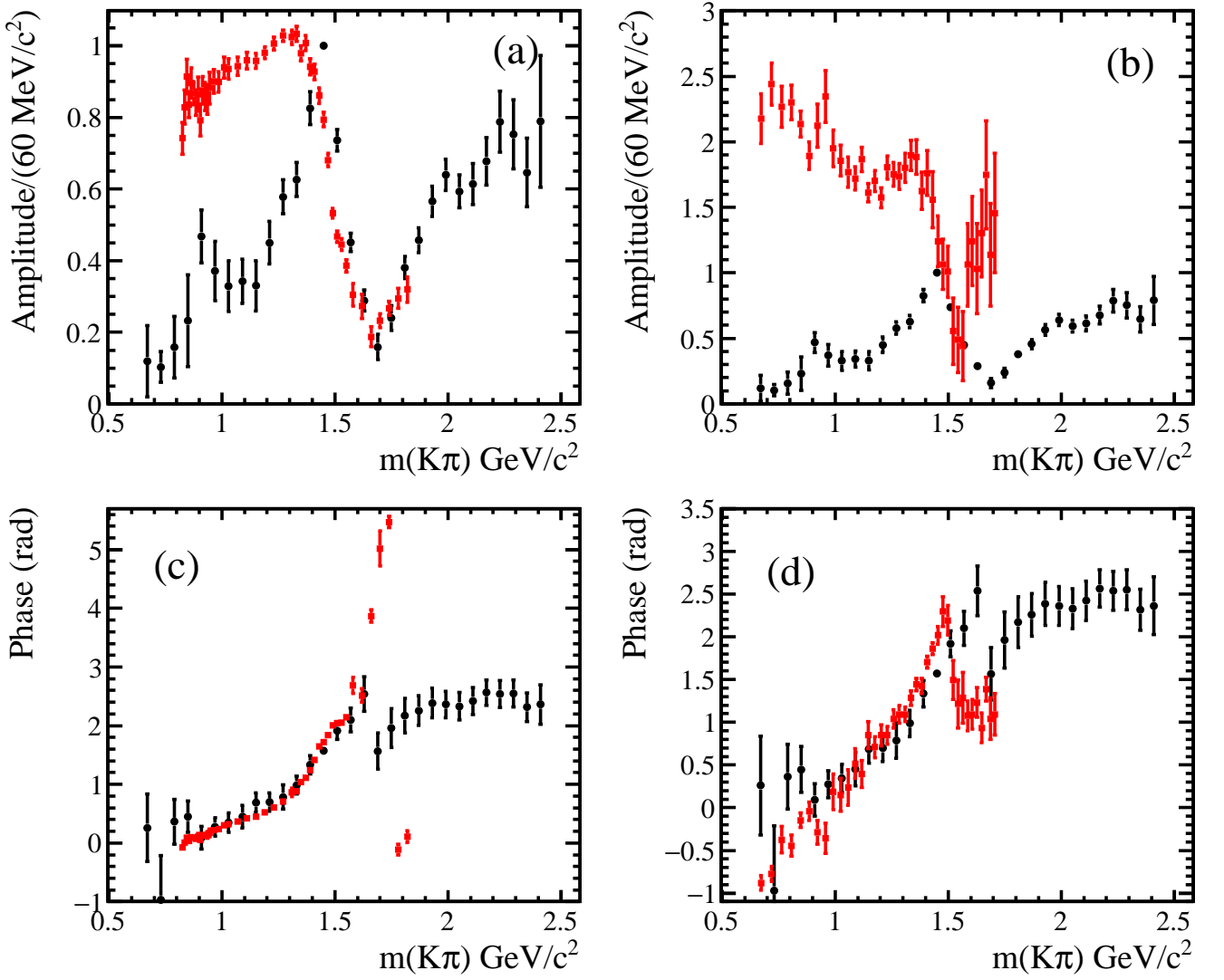}
\caption{The $I=1/2$  $K \pi$ $\mathcal{S}$-wave amplitude measurements from  $\etac \to \kskpi$ compared to the (a) LASS and (b) E791 results: the corresponding $I=1/2$  $K \pi$ $\mathcal{S}$-wave phase measurements compared to the
 (c) LASS and (d) E791 measurements. 
  Black dots indicate the results from the present analysis; square (red) points indicate the LASS or E791 results. The LASS data are plotted in the region having only one solution.}
\label{fig:fig14}
\end{center}
\end{figure*}

We compare the present measurement of the $K \pi$ $\mathcal{S}$-wave amplitude from  $\etac \to \kskpi$ with measurements
from LASS~\cite{lass_kpi} in Fig.~\ref{fig:fig14}(a)(c) and E791~\cite{aitala1} in Fig.~\ref{fig:fig14}(b)(d).
We plot only the first part of the LASS measurement since it suffers from
a two-fold ambiguity above the mass of 1.82 \gevcc.
The Dalitz plot fits extract invariant amplitudes. Consequently, in Fig.~\ref{fig:fig14}(a), the LASS $I=1/2$  $K \pi$ scattering amplitude
values have been multiplied by the factor $m(K \pi)/q$ to convert to invariant amplitude, and normalized so as to equal
the scattering amplitude at 1.5 \gevcc\ in order to facilitate comparison to the \etac results. Here $q$ is the momentum of either meson in the $K \pi$ rest frame. For better comparison, the LASS absolute phase measurements have been displaced by $-0.6$ rad before plotting them in Fig.~\ref{fig:fig14}(c).
In Fig.~\ref{fig:fig14}(b) the E791 amplitude has been obtained by multiplying the amplitude $c$ in Table III of Ref.~\cite{aitala1} by the Form Factor $F_D^0$, for which the mass-dependence is motivated by theoretical speculation. This yields amplitude values corresponding to the E791 Form Factor having value 1, as for the \etac analyses. In Fig.~\ref{fig:fig14}(d), the E791 phase measurements have been displaced by $+0.9$ rad, again in order to facilitate comparison to the \etac measurements.

While we observe similar phase behavior
among the three measurements up to about 1.5 \gevcc, we observe striking differences in the mass dependence of the amplitudes.

\section{Summary}

We perform Dalitz plot analyses, using an isobar model and a MIPWA method, of data on the decays $\eta_c \to \kskpi$ and $\etac \to \Kp \Km \piz$, where the \etac mesons are produced in two-photon interactions in the \babar\ experiment at SLAC. 
We find that, in comparison with the isobar models examined here, an improved description of the data is obtained by using a MIPWA method.

We extract the $I=1/2$  $K \pi$ $\mathcal{S}$-wave amplitude and phase and find good agreement between the measurements
for the two $\eta_c$ decay modes.
The $K \pi$ $\mathcal{S}$-wave is dominated by the presence of the $K^*_0(1430)$ resonance which is observed as a clear peak
with the corresponding increase in phase of about $\pi$ expected for a resonance. A broad structure in the 1.95 \gevcc\ mass region indicates
the presence of the $K^*_0(1950)$ resonance.

A comparison between the present measurement and previous experiments indicates a similar trend for the phase up to a mass of
1.5 \gevcc. The amplitudes, on the other hand, show very marked differences.

To fit the data we need to introduce a new $a_0(1950)$ resonance in both $\eta_c \to \kskpi$ and $\etac \to \Kp \Km \piz$ decay modes, and their associated parameter values are in good agreement. The weighted averages for
the parameter values are:

\begin{equation}
  \begin{split}
    m(a_0(1950))=1931 \pm 14 \pm 22 \ {\rm MeV}/c^2, \\
    \Gamma(a_0(1950))= 271 \pm 22 \pm 29 \ {\rm MeV}
    \end{split}
\end{equation}

\noindent with significances of 2.5$\sigma$ and 4.2$\sigma$ respectively, including systematic uncertainties.
These results are, however, systematically limited, and more detailed studies of the $I=1$ $K \Kbar$ $\mathcal{S}$-wave will be
required in order to improve the precision of these values.

\section{Acknowledgements}
We are grateful for the 
extraordinary contributions of our \pep2\ colleagues in
achieving the excellent luminosity and machine conditions
that have made this work possible.
The success of this project also relies critically on the 
expertise and dedication of the computing organizations that 
support \babar.
The collaborating institutions wish to thank 
SLAC for its support and the kind hospitality extended to them. 
This work is supported by the
US Department of Energy
and National Science Foundation, the
Natural Sciences and Engineering Research Council (Canada),
the Commissariat \`a l'Energie Atomique and
Institut National de Physique Nucl\'eaire et de Physique des Particules
(France), the
Bundesministerium f\"ur Bildung und Forschung and
Deutsche Forschungsgemeinschaft
(Germany), the
Istituto Nazionale di Fisica Nucleare (Italy),
the Foundation for Fundamental Research on Matter (The Netherlands),
the Research Council of Norway, the
Ministry of Education and Science of the Russian Federation,
Ministerio de Economia y Competitividad (Spain), and the
Science and Technology Facilities Council (United Kingdom).
Individuals have received support from 
the Marie-Curie IEF program (European Union), the A. P. Sloan Foundation (USA) 
and the Binational Science Foundation (USA-Israel).
The work of A. Palano and M. R. Pennington was supported (in part) by the U.S. Department of Energy, Office of Science, Office of Nuclear Physics under contract DE-AC05-06OR23177.

\renewcommand{\baselinestretch}{1}


\begin{thebibliography}{99}
\bibitem{conj} The use of charge conjugate reactions is implied, where not explicitly expressed, throughout the paper.
\end{thebibliography}
\end{document}